\documentclass{ptephy_v1}
\preprintnumber{UT-16-14} 

\usepackage{amsfonts}
\usepackage{amsmath}
\usepackage{amssymb}
\usepackage{color}                    
\newcommand{\ba}{\begin{eqnarray}}
\newcommand{\ea}{\end{eqnarray}}

\begin{document}

\title{Nambu bracket and M-theory}


\author{Pei-Ming Ho}
\affil{Department of Physics and Center for Theoretical Sciences,
Center for Advanced Study in Theoretical Sciences,
National Taiwan University, Taipei 106, Taiwan, R.O.C. \email{pmho@phys.ntu.edu.tw}}

\author[2]{Yutaka Matsuo}
\affil{Department of Physics, The University of Tokyo, Hongo 7-3-1, Bunkyo-ku, Tokyo 113-0033, Japan\email{matsuo@phys.s.u-tokyo.ac.jp}}

%


\begin{abstract}%
Nambu proposed an extension of dynamical system through the introduction
of a new bracket (Nambu bracket) in 1973.
This article is a short review of the developments after
his paper.  Some emphasis are put on a viewpoint that
the Nambu bracket naturally describes extended objects which appear in M-theory
and the fluid dynamics.  The latter part of the paper is devoted to a
review of the studies on the Nambu bracket (Lie 3-algebra)
in Bagger-Lambert-Gustavsson theory 
of multiple M2-branes.
This paper is a contribution to the proceedings of Nambu memorial symposium
(Osaka City University, September 29, 2015). 
\end{abstract}

\subjectindex{xxxx, xxx}

\maketitle

\section{Introduction}
Nambu's contributions to Physics are profound
and diverse.  While creating great ideas such as 
spontaneous symmetry breaking
which becomes standard in the contemporary Physics,
he sometimes presented ideas which were mysterious 
in the beginning but became gradually recognized after years.
Nambu bracket \cite{Nambu:1973qe} may be one of latter examples.
The importance might not be so obvious even for himself.
According to the paper, he kept the idea for
more than twenty years before the publication.
If we take it as was written, it started in early 50s
when he moved from Osaka City University to Princeton.
The reason why he needed so long period to decide the publication
is understandable from his paper.
Just after the definition of the bracket, he pointed out serious
obstacles for his generalized dynamical system. During the long
period that he kept his idea, he developed 
various new ideas which are useful and stimulating
even from the current viewpoints.

As described in \cite{Nambu:1973qe}, there are two major challenges in the subject.
One is how to quantize the Nambu bracket and the other is
multi-variable extensions.  This turned out to be difficult or impossible
(there appeared the no-go theorems).
We have to relax  ``natural" requirements of the Nambu bracket
which are the direct generalization of the Poisson bracket.
The ways to relax the conditions are not unique and
depend on the problem we are considering.  It explains
the existence of many proposals to define (quantum) Nambu bracket.

The purpose of this article is to give a
brief review of the Nambu bracket and
to illuminate some applications in M-theory.
In section 2, we explain the basic material in
the original paper \cite{Nambu:1973qe}
where many ideas were already written. 
We also briefly quote some of the important results
since then.  It turned out that Nambu bracket
fits with M-theory well and there appeared varieties of applications. 
We put some emphasis on the matrix model description of M-theory.
In section 3, we review a proposal by Takhtajan \cite{Takhtajan:1993vr}
that the Nambu bracket naturally describes the
extended object. For the 3-bracket case, it corresponds to strings.
In this respect, it fits non-canonical string such as the self-dual
string on M5-brane and the vortex in the incompressible
fluid.  We explain the quantization of Takhtajan's action
which might be relevant to describe these non-canonical strings.
Finally in section 4, we review the developments of the Nambu bracket
and associated Fillipov Lie 3-algebras
to describe the multiple
M2-branes by Bagger, Lambert and Gustavsson (BLG model)
\cite{Bagger:2006sk,Bagger:2007jr, Bagger:2007vi, Gustavsson}.
Special emphasis is put on our works where we introduced
varieties of Lie 3-algebras with
Lorentzian signature in BLG formalism to describe different types of 
extended objects appearing in M-theory 
and string theory.

\section{Nambu bracket}
\subsection{An introduction of Nambu bracket}
In 1973 \cite{Nambu:1973qe},  Nambu proposed a generalization of Poisson bracket
defined on a canonical pair $x,p$
\begin{eqnarray}
\left\{ f, g\right\}=\frac{\partial f}{\partial x}\frac{\partial g}{\partial p}-
\frac{\partial f}{\partial p}\frac{\partial g}{\partial x}\,,
\end{eqnarray}
by the introduction of new dynamical system 
based on a canonical {\em triple} $x_1, x_2, x_3$:
\begin{eqnarray}\label{NB1}
\{f,g,h\}=\sum_{ijk}\epsilon_{ijk}\frac{\partial f}{\partial x_i}\frac{\partial g}{\partial x_j}
\frac{\partial h}{\partial x_k}=:
\frac{\partial(f,g, h)}{\partial(x_1, x_2, x_3)}
\,.
\end{eqnarray}
This bracket was later referred to as the Nambu bracket.
Instead of the canonical Hamiltonian equation,
\begin{eqnarray}\label{PD}
\dot f=\left\{f, H\right\}\,,
\end{eqnarray}
the time evolution is defined by
the new bracket with two Hamiltonians $H, G$,
\begin{eqnarray}\label{ND}
\dot f=\{f, H, G\}\,.
\end{eqnarray}
As the Hamiltonian is a constant of the motion in (\ref{PD}),
two Hamiltonians $H, G$ are constant of the motion under the Nambu dynamics
(\ref{ND}),
\begin{eqnarray}
\dot H=\{H, H, G\}=0,\quad
\dot G=\{G, H, G\}=0\,,
\end{eqnarray}
due to the antisymmetry of the bracket.

Just as the canonical Hamiltonian equation (\ref{PD}) keeps the
infinitesimal area of the phase space, $\Delta x \Delta p$, the generalized
system (\ref{ND}) keeps the volume of the triple $\Delta x_1\Delta x_2\Delta x_3$:
\begin{eqnarray}
\vec \nabla \cdot \vec{v}=0,\quad
\vec{v}:=\dot{\vec{x}}=\{\vec x, H, G\}=\vec\nabla H\times \vec\nabla G\,.
\end{eqnarray}
In this sense, it defines a dynamical system which has a 
generalized Liouville property
(conservation of phase volume).
This was one of the reasons why Nambu introduced such bracket.

As an example which is described by the new bracket, Nambu
considered the rotational motion of a rigid body which is described
by angular momentum $J_x, J_y, J_z$.
In this case, we have two conserved quantities,
the energy and the total momentum:
\begin{eqnarray}
H=\frac{J_x^2}{2I_x}+\frac{J_y^2}{2I_y} +\frac{J_z^2}{2I_z}\,,\quad
G=\frac{J_x^2+J_y^2+J_z^2}{2}=\frac{\mathbf{J}^2}{2}\,,
\end{eqnarray}
where $I_x, I_y, I_z$ are the moment of inertial along each axis.
We introduce the Nambu bracket by
\begin{eqnarray}
\{f, g, h\}=\frac{\partial(f,g,h)}{\partial(J_x, J_y, J_z)}\,.
\end{eqnarray}
Some computation shows that the equation (\ref{ND})  gives
Euler's equation for the rigid body:
\begin{eqnarray}
\dot J_x=\left(\frac{1}{I_y}-\frac{1}{I_z}\right) J_y J_z,\quad
\dot J_y=\left(\frac{1}{I_z}-\frac{1}{I_x}\right) J_z J_x,\quad
\dot J_z=\left(\frac{1}{I_x}-\frac{1}{I_y}\right) J_x J_y\,.
\end{eqnarray}

\subsection{Generalizations of Nambu bracket}
\subsubsection{Mathematical definition}
Nambu bracket is defined in more abstractly 
through the following requirements which generalize
those for the Poisson bracket.
It is defined on the ring of $C^\infty$ functions $\mathcal{A}$
with $M$ variables $x_1,\cdots, x_M$. The Nambu bracket
in a generalized sense is defined by a map
$\mathcal{A}^{\otimes N}\rightarrow \mathcal{A}$
\begin{eqnarray}
f_1,\cdots,f_N \in \mathcal{A}\Rightarrow \{f_1,\cdots,f_N\}\in \mathcal{A}
\end{eqnarray}
which satisfies the following three conditions \cite{Takhtajan:1993vr}:
\begin{itemize}
	\item[I)] Alternation law (skew symmetry):
	\begin{eqnarray}\label{alternation}
	\{f_{\sigma(1)},\cdots, f_{\sigma(N)}\}=(-1)^\sigma \{f_1,\cdots,f_N\}\,\quad\mbox{for arbitrary } \sigma\in\mathfrak{S}_N\,.
	\end{eqnarray}
	\item[II)] Derivative law (Leibniz rule):
	\begin{eqnarray}\label{derivative}
	\{fg,f_2,\cdots,f_N\}=f\{g,f_2,\cdots,f_N\}+g\{f,f_2,\cdots, f_N\}\,.
	\end{eqnarray}
	\item[III)] Generalized Jacobi law (fundamental identity):
	\begin{eqnarray}\label{Fundamental}
	\{\{f_1,\cdots, f_{N}\},g_1,\cdots, g_{N-1}\}=\sum_{i=1}^N
	\{f_1,\cdots \{f_i, g_1,\cdots, g_{N-1}\},\cdots, f_N\}\,.
	\end{eqnarray}
\end{itemize}
These rules are essential to define the time evolution
of Nambu equation with $N-1$ Hamiltonians:
\begin{eqnarray}\label{higherNambu}
\frac{df}{dt}=\{f,H_1,\cdots, H_{N-1}\}\,.
\end{eqnarray}
or a canonical transformation of variables defined by
generating functions $S_1,\cdots, S_{N-1}$
(for $N=M$):
\begin{eqnarray}\label{hC}
\delta x_i=\{x_i,S_1,\cdots, S_{N-1}\}\,.
\end{eqnarray}
They are natural in the sense to 
ensure the basic properties of
the dynamics.
Firstly, the alternation law I) ensures the Hamiltonians
are constants of the motion\footnote{It implies that the Nambu dynamical system
has higher conserved quantities $H_1,\cdots, H_{N-1}$.
In this sense, it has some connection with the integrable models.
See, for example \cite{Curtright:2002sr}, for a study
in this direction.
}:
\begin{eqnarray}\label{higher_conservation}
\frac{dH_i}{dt}=\{H_i,H_1,\cdots, H_{N-1}\}=0.
\end{eqnarray}
The derivative law II) implies Leibniz rule for
the time derivative:
\begin{eqnarray}
\frac{d(fg)}{dt}&=&\{fg,H_1,\cdots, H_{N-1}\}\nonumber\\
&=& f\{g,H_1,\cdots,H_{N-1}\}+\{f,H_1,\cdots, H_{N-1}\}g=\frac{df}{dt}g+f\frac{dg}{dt}\,.
\end{eqnarray}
Finally the fundamental identity  III) 
(in the following we abbreviate it as FI)
 implies the distribution law of the time derivative
in the bracket:
\begin{eqnarray}
\frac{d}{dt}\{f_1,\cdots, f_N\}=\sum_{i=1}^N \{f_1,\cdots, \frac{df_i}{dt},\cdots, f_N\}\,.
\end{eqnarray}

\subsubsection{Some properties of the generalized Nambu bracket}
Here is a few comments on the generalized Nambu bracket
and Liouville theorem:
\begin{itemize}
	\item The Jacobian \cite{Nambu:1973qe}
	\begin{eqnarray}\label{higher}
	\{f_1,\cdots, f_n\}:=\frac{\partial(f_1,\cdots,f_n)}{\partial(
		x_1,\cdots, x_n)}
	\end{eqnarray}
	satisfies all conditions I)--III) for $N=M=n$.  The time evolution
	defined by this bracket keeps the
	$n$-dimensional phase volume
	$\Delta x_1\cdots \Delta x_n$,
	thus the dynamics satisfies the Liouville theorem.
\item 
	In \cite{Takhtajan:1993vr}, possible
	solutions to the conditions I) II) III) are examined.
	The bracket which satisfies I) and II) may be written in the form:
	\begin{eqnarray}
	\{f_1,\cdots, f_N\}=\sum_{i_1,\cdots, i_N} \eta_{i_1\cdots i_N}(x)
	\partial_{i_1} f_1\cdots \partial_{i_N} f_N
	\end{eqnarray}
	where $\eta_{i_1\cdots i_N}$ is anti-symmetric for the indices.
	The fundamental identity is written as the bilinear identities
	among Nambu tensor $\eta_{i_1\cdots i_N}(x)$.  It was
	proved that Nambu bracket should be decomposable 
\begin{eqnarray}
\eta:=\eta_{i_1\cdots i_N}\partial_{i_1}\wedge\cdots \wedge\partial_{i_N}
=V_1\wedge\cdots \wedge V_N,\quad V_a=\sum_i v_a^i\partial_{x_i}
\end{eqnarray}	
	to satisfy the constraint \cite{gautheron1996some}. In particular,
	a natural multi-variable extension such as $\eta=\partial_1\wedge\partial_2\wedge \partial_3+\partial_4\wedge\partial_5\wedge \partial_6$ does not satisfy FI.
	\item In order to keep the phase volume, it is possible to 
	generalize (\ref{higherNambu}) to 
	\begin{eqnarray}\label{general_Nambu}
	\frac{df}{dt}=\sum_{\alpha=1}^Q \{f, H_1^{(\alpha)},\cdots, H_{N-1}^{(\alpha)}\}\,,
	\end{eqnarray}
	with $Q(N-1)$ Hamiltonians $H_i^{(\alpha)}$.  
	These generalized Hamiltonians, however,
	are not preserved by the equation of motion.
	In terms of the canonical variables, the equation of motion
	is written as
	\begin{eqnarray}
	\dot{x}_i=\sum_{j=1}^N \partial_j f_{ij}(x),\quad
	f_{ij}:=\sum_{k_1,\cdots, k_{N-2}} \epsilon_{ijk_1\cdots k_{N-1}}\sum_{\alpha}^{Q}
	H_1^{(\alpha)}\frac{\partial(H_2^{(\alpha)},\cdots, H_{N-1}^{(\alpha)})}{\partial(x_{k_1},\cdots, x_{k_{N-2}})}\,.
	\end{eqnarray}
	The quantity $f_{ij}$ is antisymmetric $f_{ij}=-f_{ji}$.
	The first equation is the most general form to preserve phase volume.
	\item For $N=3$ case, the canonical equation is
	 rewritten as
	\begin{eqnarray}\label{eom}
	\dot{\vec{x}}=\vec\nabla\times \vec{A}, \quad \vec{A}=\sum_{\alpha=1} H_\alpha
	\vec\nabla G_\alpha\,.
	\end{eqnarray}
	It was noted \cite{Nambu:1973qe} that there are some arbitrariness in the choice of
	$H_\alpha, G_\alpha$ to give the same equation.  
	Namely a different set $H'_\alpha, G'_\alpha$ of Hamiltonian
	gives the same equation of motion as long as it satisfies
	canonical transformation with $(H_\alpha, G_\alpha)$ as the canonical pair
	in the Poisson sense,
	\begin{eqnarray}
	[H'_\alpha, G'_\beta]:=\sum_{\gamma=1}^N \frac{\partial(H'_\alpha,G'_\beta)}{\partial(H_\gamma,G_\gamma)}
	=\delta_{\alpha\beta},\quad [H'_\alpha, H'_\beta]=[G'_\alpha, G'_\beta]=0\,.
	\end{eqnarray}
	One may check the statement for the infinitesimal variations.
	Let us use $\delta H_\alpha=H'_\alpha-H_\alpha=\epsilon\frac{\partial S(H,G)
	}{\partial G_\alpha}$ and $\delta G_\alpha=G'_\alpha-G_\alpha=-\epsilon\frac{\partial S(H,G)
}{\partial H_\alpha}$. The variation of the equation (\ref{eom}) 
is absorbed in the variation of $\vec A$ as,
$
\delta \vec{A} = \epsilon\vec\nabla\left(S-\sum_\alpha H_\alpha\frac{\partial S}{
	\partial H_\alpha}\right)
$
which may be interpreted as the infinitesimal gauge transformation.
It is obvious that it leads to the same equation of motion.

\item The other type of the hierarchy structure
exists for general $n$ \cite{Takhtajan:1993vr}.
Starting from arbitrary $n+1$ bracket 
$\{f_1,\cdots, f_{n+1}\}$
which satisfies I)-III), one may define
the $n$ bracket by using arbitrary $K$,
\begin{eqnarray}
\{f_1,\cdots, f_n\}_K:=\{f_1,\cdots, f_n, K\}\,.
\end{eqnarray} 
One may show easily that the new bracket
satisfies the three conditions. By continuing the same procedure,
one may obtain Nambu $m$ bracket from Nambu $n$ bracket
for $m<n$.

As an example, let us take the Nambu bracket for the
rigid rotor.  The original Nambu bracket was
defined as
\begin{eqnarray}
\{f,g, h\}=\frac{\partial(f,g,h)}{\partial(J_x,J_y, J_z)}\,.
\end{eqnarray}
If we take $K=\frac{1}{2}(J_x^2+J_y^2+J_z^2)$,  
the Poisson bracket $\{\bullet,\bullet\}_K:=\{\bullet,\bullet,K\}$ gives
\begin{eqnarray}
\{J_x, J_y\}_K= J_z,\quad
\{J_y, J_z\}_K= J_x,\quad
\{J_z, J_x\}_K= J_y,
\end{eqnarray}
which is the standard Poisson bracket for the
angular momentum.
\end{itemize}

\subsection{Difficulties in Nambu bracket}
In \cite{Nambu:1973qe},
it was already mentioned some serious difficulties in
the formulation. 
They are not the technical problems and there is no way
to overcome them.  All we can do is to
relax some of the conditions I), II), III)
as long as they do not produce serious troubles
in the applications which we consider.

\paragraph{Multi-variable extension}
In Poisson bracket, it is straightforward to 
extend the formalism to $2N$ canonical pairs,
$x^i, p_i$
($i,j=1,\cdots, N$) as
\begin{eqnarray}
\{f,g\}=
\sum_{i=1}^N \left(\frac{\partial f}{\partial x_j}\frac{\partial g}{\partial p^j}-\frac{\partial f}{\partial p^j}\frac{\partial g}{\partial x_j}\right)
\end{eqnarray}
It satisfies the consistency condition of the Poisson bracket
(Jacobi identity),
\begin{eqnarray}\label{Jacobi}
\{\{f,g\},h\}+\{\{g,h\},f\}+\{\{h,f\},g\}=0\,,
\end{eqnarray}
for any $N$.  
The existence of such identity is necessary
for the compatibility of the time evolution
(\ref{PD}). 
%
%

In the Nambu bracket, the analog of (\ref{Jacobi}) is played
by the fundamental identity (FI).
A difficulty of the Nambu bracket is that the FI is too strict
that there is almost no room for the generalization.  
As already mentioned,
a naive multi-variable extension of (\ref{NB1})
\begin{eqnarray}\label{mNambu}
\{f,g,h\}=\sum_{a=1}^N
\frac{\partial(f,g, h)}{\partial(x^a_1, x^a_2, x^a_3)}
\end{eqnarray}
for $3N$ variables $x^a_i$ ($a=1,\cdots, N$, $i=1,2,3$) does not satisfy FI.
In \cite{Nambu:1973qe}, Nambu examined the canonical transformation
defined by the bracket (\ref{mNambu}) and the generating function
$S_i$ in (\ref{hC}) should be decomposed as $S_i=\sum_a S_i^a(x^a)$
from the consistency conditions. It implies that the variable set
$(x_1^a,x_2^a,x_3^a)$ should transform within themselves.
While the fundamental identity was not
proposed explicitly but this analysis has already shown
the difficulty in the multi-variable extension.

\paragraph{Quantization}
In the Poisson bracket, the quantization procedure
is to replace the bracket into the commutator
\begin{eqnarray}
\{f,g\}=\frac{\partial(f,g)}{\partial(x,p)} \rightarrow 
\left[\hat f, \hat g\right]=\hat{f}\hat{g}-\hat{g}\hat{h}\,.
\end{eqnarray}
The commutator satisfies a noncommutative version of
the three consistency conditions.

For the Nambu bracket, the most straightforward
generalization of the commutator is,
\begin{equation}\label{canonical triple}
[X,Y,Z]=XYZ+YZX+ZXY-YXZ-XZY-ZYX=X[Y,Z]+Y[Z,X]+Z[X, Y]\,.
\end{equation}
While it satisfies I), the conditions II) and III)
are not kept.
%

\paragraph{Solutions to canonical quantization condition}
While it does not satisfy the conditions, it may be possible
to use it relaxing some conditions.
In \cite{Nambu:1973qe}, Nambu tried to find a set of operators 
which satisfies an analog of canonical quantization condition:
\begin{eqnarray}\label{canN}
[X_a, Y_b, Z_c]=i\delta_{abc}
\end{eqnarray}
while neglecting the constraints (2,3) for the moment.
Here $\delta_{abc}=1$ when $a=b=c$ and $=0$ otherwise.
Assuming the set $\{X_a, Y_a, Z_a\}$ ($a=1,\cdots, N$) 
are the basis of some Lie algebra $\mathfrak{g}$.
By writing
\begin{eqnarray}
[X_1, Y_1]=i Z', \quad [Y_1, Z_1]=i X',\quad [Z_1, X_1]=i Y'
\end{eqnarray} 
for the first three generators and $X', Y', Z'\in \mathfrak{g}$.
Eq.(\ref{canN}) implies that
\begin{eqnarray}
X_1 X'+Y_1 Y' +Z_1 Z'=1\,.
\end{eqnarray}
The right hand side is c-number and should commute with
arbitrary generators in $\mathfrak{g}$.  So it may be implemented by
Casimir operator for the Lie algebra.  From this observation, assuming
$\mathfrak{g}$ is semisimple, one may classify the possible algebras.
The result is:
\begin{eqnarray}
SO(3), SO(2,1), SO(4), SO(3,1)\,.
\end{eqnarray}
If the algebra is not semi-simple, 
there are futher choices after contractions:
\begin{eqnarray}
E(3), E(2,1), E(2), E(1,1)
\end{eqnarray}
Here $E(3)$ is the euclidean algebra generated by $\vec P, \vec L$
(momentum and angular momentum operators).
The others are similar algebra with different dimensions and signature.

\paragraph{Use of nonassociative algebras}
Nambu also considered a possibility to use nonassociative
algebra to define the quantization.
In this case, the associator
\begin{eqnarray}
(a,b,c)=(ab)c-a(bc)
\end{eqnarray}
does not in general vanish.  If we require that the
associator be skew symmetric with respect to all elements,
the algebra is restricted to Cayley number.
It nevertheless does not satisfy the derivative property.

He then modified the bracket to keep the derivative property:
\begin{eqnarray}
D(a,b;x)=D(a,b)x:= a(bx)-b(ax)+(xb)a-(xa)b+(bx)a-b(xa)
\end{eqnarray}
for the Cayley number.  This time, we do not have total skewness
but only the partial one:
$D(a,b;x)=-D(b,a;x)$.
The time evolution generated by
\begin{eqnarray}
\frac{dx}{dt}=\sum_i D(H_i,G_i)x
\end{eqnarray}
generates the $G_2$ automorphism.\footnote{It looks like
Nahm equation with $G_2$ holonomy
if $H$ and $G$ are properly chosen.
It may provide another link with M-theory.
See for example, \cite{Cherkis:2014xua}.}

He also examined to use a commutative and nonassociative
algebra (Jordan algebra).  In this case, the derivative
operator is written in the form:
\begin{eqnarray}
D(a,b)x=(a,b,x)-(b,a,x)\,.
\end{eqnarray}
Jordan algebra, in general, is written in terms of 
noncommutative and associative algebra by modification of
the multiplication $a\cdot b=(ab+ba)/2$.  If we use
this realization, the derivative operator is rewritten
as $D(a,b)x=[x,[a,b]]$. So the equation of motion
is reduced to the conventional Hamiltonian flow where
Hamiltonian is written in the form $[H,G]$.


\subsection{Some attempts to quantize Nambu bracket}

A natural approach to quantize the Nambu bracket
is through the deformation quantization.
It is a generalization of Moyal
bracket,
\begin{eqnarray}
&& f\star g:=e^{\frac12\hbar \epsilon_{ij} \partial^{(1)}_i \partial^{(2)}_j}f(x^{(1)})
g(x^{(2)})|_{x^{(1)}=x^{(2)}=x}\nonumber\\
&&\quad \rightarrow
(f, g,h):=e^{\frac16 \hbar \epsilon_{ijk} \partial^{(1)}_i \partial^{(2)}_j\partial^{(3)}_k}f(x^{(1)})
g(x^{(2)})h(x^{(3)})|_{x^{(1)}=x^{(2)}=x^{(3)}=x}\,.
\end{eqnarray}
The quantum Nambu bracket thus defined failed to
satisfy FI \cite{Takhtajan:1993vr}.
There are a few alternative approaches for the deformation quantization
(see for example, \cite{gautheron1996some, Curtright:2002sr}).
Later, Dito et. al. \cite{Dito:1996xr}
proposed a deformation quantization based on Zariski quantization
which satisfies FI.  It is very different from
conventional quantization method but some efforts have
been made to use it for the M-theory \cite{Minic:1999js}.

Curtight and Zachos tried to formulate the quantum Nambu bracket
in the line of (\ref{canonical triple}). Instead of the modification
of the bracket (\ref{canonical triple}), they proposed an alternative to
the fundamental identity \cite{Curtright:2002fd}. This reference contains
a nice review on the Nambu bracket.

In the connection with the matrix model approach to M-theory 
\cite{Banks:1996vh}, the Nambu dynamics
is natural to realize the generalized uncertainty relation
$\Delta p\Delta q\Delta r\geq \hbar$.
Awata, Li, Minic and Yoneya \cite{Awata:1999dz}
defined a quantization of Nambu bracket through the matrices as
\begin{eqnarray}
[A,B,C]:=\mbox{Tr}(A)[B,C]+\mbox{Tr}(B)[C,A]+\mbox{Tr}(C)[A,B]\,,
\end{eqnarray}
which satisfies the fundamental identity.
Very recently, Yoneya suggested a similar bracket \citep{Yoneya:2016wqw}
to describe the covariant M-theory matrix model.

In the context of M-theory, the degree of
freedom is predicted to behave as $O(N^3)$ for $N$ five-branes
from AdS/CFT correspondence. In this sense, it may be natural
that the quantum degree of freedom is described by a tensor
with three indices $A_{ijk}$ (cubic matrix).
Such direction was pursued by Kawamura
in \cite{Kawamura:2002yz, Kawamura:2003cw}.
The triple matrix for the cubic matrix was defined as
\begin{eqnarray}
(ABC)_{lmn}=\sum_k A_{lmk}B_{lkn}C_{kmn}\,,
\end{eqnarray}
and quantum Nambu bracket is defined by anti-symmetrization.
While FI is not satisfied with this bracket,
a consistent dynamical system can be constructed
if the Hamiltonians are restricted to the normal form,
\begin{eqnarray}
H_{lmn}=\delta_{lm} h_{mn}+\delta_{nm}h_{ln}+\delta_{ln}h_{ml}.
\end{eqnarray}
Due to this restriction, the time evolution becomes essentially
diagonal.
We note that the choice of the product of the cubic matrix is
not unique.  For example, in \cite{Ho:2007vk}, a different choice,
$(ABC)_{lmn}=\sum_{ijk} A_{ij n}B_{jkl}C_{kim}$ was used.
It is more natural to associate the cubic matrix with
the triangle which covers the membrane: the index is
assigned to the edges of a triangle and the triple product
is interpreted as gluing edges of three triangles to produce
three open edges.  It is a natural framework
to implement discretized quantum gravity \cite{Turaev:1992hq}
but the analog of FI is difficult to be realized.

\section{Nambu bracket and the extended objects}

\subsection{Takhtajan's action}
In \cite{Takhtajan:1993vr}, Takhtajan introduced an action principle
which describes the Nambu dynamics as the motion of the extended objects.
Let new variables $X^i(\sigma, t)$ ($i=1,2,3$) describe a string-like object
in $\mathbb{R}^3$ (three spacial dimensions).  We assume that the Hamiltonians $H, K$ are the
functions of $X^i(\sigma, t)$ at the same world-sheet point.
\begin{eqnarray}\label{Tk_action}
S= \frac{1}{3}\int dtd\sigma \epsilon_{ijk} X^i\partial_t X^j \partial_\sigma X^k+
\int dtd\sigma H\partial_\sigma K\,.
\end{eqnarray}
Variation of the action gives
\begin{eqnarray}\label{varTA}
\delta S=\int dtd\sigma 
\left(\frac{1}{3}\epsilon_{ijk}\partial_t X^i- \frac{\partial(H,K)}{\partial(X^j,X^k)}\right)
\frac{\partial X^j}{\partial \sigma} \delta X^k
\end{eqnarray}
It implies the equation of motion for the string-like object,
\begin{eqnarray}
\partial_t X^i-\frac12 \epsilon_{ijk}\frac{\partial(H,K)}{\partial(X^j,X^k)}\propto
\partial_\sigma X^i\,.
\end{eqnarray}
The left hand side of the equation is Nambu's equation and the right hand side
is the arbitrariness due to the reprametrization invariance with respect to $\sigma$.
When we need to consider more general Nambu action of the form (\ref{general_Nambu}),
one may simply replace it by 
\begin{eqnarray}
S=\int dt d^{N-2}\sigma
\left( \frac{1}{N!}
\epsilon_{i_1,\cdots, i_N}
X^{i_1}\frac{\partial(X^{i_2}\cdots  X^{i_{N}})}{\partial(t,
\sigma_1,\cdots,\sigma_{N-2})}
-　H_1 \sum_\alpha　\frac{\partial(H^\alpha_2,\cdots, H^\alpha_{N-1})}{\partial(\sigma_1,\cdots,\sigma_{N-2})} \right)
\end{eqnarray}
In this case, the variable $X^i(\sigma, t)$ describes an $(N-2)$-brane.

Takhtajan's action is relevant to the study of self-dual string
on M5-brane \cite{Bergshoeff:2000jn, Kawamoto:2000zt} and 
the fluid motion in 3 dimensions.  The connection with the fluid
motion is discussed in the next subsection.
In the context of M-theory, the fundamental
degree of freedom is described by M2-brane (and the dual M5-brane)
whereas the effective description by supergravity is
described by anti-symmetric 
3-form field $C$
and its dual 
6-form.
In the low energy, the effective description of the membrane
is given by Nambu-Goto type action and the coupling to
three-form $C$,
\begin{eqnarray}
S=\int d^3\sigma T \det\left(-G\right)+\int_V C,
\end{eqnarray}
where $T$ is the membrane tension and $V$ is the world
volume of the membrane.
Suppose we are considering an extreme situation
where 
$C$ is constant
and 
large enough such that one may neglect the Nambu-Goto part,
we are left with the coupling of the membrane world-volume to the constant
3-form field.  In the simplest case where $C_{012}\neq 0$, the latter
term coincides with the Takhtajan action when the world-volume
has the boundary since
\begin{eqnarray}
\frac{1}{3!}C_{012}\int_V \epsilon_{ijk} dX^i\wedge dX^j\wedge dX^k=\frac{1}{3!} C_{012} \int_{\partial V} \epsilon_{ijk}X^i dX^j\wedge dX^k.
\end{eqnarray}
It is known that the the boundary of M2-brane is located
on M5-brane. On M5-brane, the two-form gauge
field should be self dual, namely $C=\star C$.  In this sense,
Takhtajan string describes the self-dual string on M5. 

\subsection{Connections with incompressible fluid dynamics}
As Nambu himself pursued for a long time, 
(due to a review in \cite{Saitou:2014vwa}),
the Nambu dynamics is a natural framework to
describe the incompressible fluid motion.
The incompressibility implies that the volume element
$\Delta v$ does not change in the time evolution.
It implies that the coordinates $\vec x(\vec x_0,t)$
has to satisfy $\frac{\partial(\vec x)}{\partial(\vec x_0)}=1$
in the Lagrangian formulation where $\vec x(\vec x_0,t)$ is the
location of fluid which was at $\vec x_0$ at $t=t_0$.
It implies that the time evolution should be written in the form,
\begin{eqnarray}
\partial_t \vec x(\vec x_0, t)=\sum_\alpha \left\{\vec x,H_\alpha(\vec x_0,t), K_\alpha(\vec x_0,t)\right\}.
\end{eqnarray}
In this subsection, we collect some descriptions of fluid motion
by the Nambu-bracket.

\subsubsection{Vortex string dynamics}
Takhtajan's action for the Nambu dynamics can be directly related with the
vortex motion where there is no dissipation. 
In the following, we use the description
in \cite{Lund:1976ze, Matsuo:1993ie}.   
We consider the Euler equation,
\begin{eqnarray}
\frac{\partial V^i}{\partial t} = V^j\partial^i V_j -V^j\partial_j V^i 
\end{eqnarray}
for the velocity $\vec V(z)$.  
In such a system, the fluid motion is governed by the center of vorticity,
described by strings localized at 
$\vec x=\vec X_I(\sigma_I,\tau)$.  
As long as there is no dissipation, the delta-function shape
vorticity retains its form and motion of the vortex
string determines the flow.
Here we assume there are $N$ vortex filaments and $I=1,\cdots, N$.
The vorticity is described by
\begin{eqnarray}
\vec \omega(x)&=& \vec{\nabla} \times \vec V= \sum_{I=1}^N\Gamma_I\int d\sigma_I
\frac{\partial \vec X_I(\sigma_i,t)}{\partial \sigma_I}\delta^{(3)}(\vec x-\vec X_I(\sigma_I, t))\,.
\end{eqnarray}
From this expression, one obtains the velocity field by Biot-Savart law,
\begin{eqnarray}
\vec V(x) &=& \sum_{I} \frac{\Gamma_I}{4\pi}\int d\sigma_I \frac{\partial \vec X_I}{\partial \sigma_I}
\times\frac{\vec x-\vec X_I}{|\vec x-\vec X_I|^3}
=\vec{\nabla}\times\sum_I \frac{\Gamma_I}{4\pi} \int d\sigma_I \frac{\partial \vec X_I}{\partial \sigma_I} \frac{1}{|x-\vec X_I|}\,.
\end{eqnarray}
Plug it into the Euler equation for the vorticity,
\begin{eqnarray}
\frac{\partial\vec \omega}{\partial t}= -\nabla\times (\vec{\omega}\times \vec V),
\end{eqnarray}
one finds that the Euler equation is solved if $\vec X_I$ satisfies
the equation,
\begin{eqnarray}
\frac{\partial \vec X_I}{\partial \sigma_I}
\times \frac{\partial \vec X_I}{\partial t}=
\frac{\partial \vec X_I}{\partial \sigma_I}\times \vec V(X_I(\sigma_I, t))\,.
\end{eqnarray}
It implies that $\frac{\partial \vec X_I}{\partial t}=\vec V(X_I(\sigma_I, t))
+\alpha \frac{\partial \vec X_I}{\partial \sigma_I}$, namely
the velocity of the string is identical to the flow velocity up to
reparametrization. The fact that the above equation 
takes the same form as 
(\ref{varTA}) implies that
the action can be written in the Takhtajan form:
\begin{eqnarray}
S&=& \int dt (L_0-E)\,,\\
L_0&=& \sum_{I=1}^N \frac{\Gamma_I}{3!}\int d\sigma_I \vec X_I\cdot
\frac{\partial\vec X_I}{\partial \sigma_I}\times \frac{\partial \vec X_I}{\partial t}\,,\\
E&=& \frac12 \int d^3 x |\vec V(x)|^2=\frac{1}{8\pi}\sum_{IJ} 
\int d\sigma_I d\sigma_J \Gamma_I\Gamma_J
\left(\frac{\partial \vec X_I}{\partial \sigma_I}\cdot
\frac{\partial \vec X_J}{\partial \sigma_J}\right) \frac{1}{|\vec X_I-\vec X_j|}\,.
\end{eqnarray}
The second term  may be rewritten as
\begin{eqnarray}
\sum_I \int d\sigma  \Gamma_I \vec{U}(\vec X_I)\cdot\frac{\partial\vec X_I}{\partial \sigma},\quad\mbox{where }
\vec U(x)=\sum_{J}\frac{\Gamma_J}{8\pi }
\int d\sigma_J \frac{\partial \vec X_J}{\partial \sigma}\frac{1}{|\vec x-\vec X_J(\sigma)|}\,.
\end{eqnarray}
One may regard it as a generalization of Takhtajan action with
the  Hamiltonians replaced by
$H^i_I=\Gamma_I X^i_I, K^i_I=U^i(\vec X_I)$ with $\alpha$ replaced by multiple indices $i,I$.

\subsubsection{Fluid dynamics in shallow water}
More recently, a totally different way of rewriting 
fluid dynamics as Nambu equation was developed in \cite{salmon2007general,nevir2009energy,sommer2009conservative}.
The shallow water equation,
\begin{eqnarray}
\dot u=h\omega v-\Phi_x,\quad
\dot v=-h\omega u-\Phi_y,\quad
\dot h = (-hu)_x-(hv)_x
\end{eqnarray}
where $(u,v)$ is the velocity for horizontal directions, 
$h$ is the fluid depth, $\omega=(v_x-u_y)/h$ is the vorticity, 
and $\Phi=\frac{u^2+v^2}{2}+gh$ is the energy density.
It was realized that the equations can be written in the form of
Nambu dynamics $\dot F=\{F, H, Z\}$ where $H=\int d^2 x h\Phi(x,y)$ 
and $Z=\int d^2x h G(q(x,y))$, where $G$ is an arbitrary function.
The bracket is defined as the functional deferentiation by
$u,v,h$ which is more involved.  See for example, eq.(1.15)
in \cite{salmon2007general}.

\subsection{Quantization of Takhtajan's action}
One may apply the standard quantization method
to Takhtajan action. We refer to \cite{Matsuo:1993ie, Bergshoeff:2000jn, Kawamoto:2000zt, Pioline:2002ba} for 3-bracket cases and \cite{Matsuo:2000fh} for higher cases.

We note that in the action (\ref{Tk_action}), the time derivative is contained
in the first term.  The momentum variable is therefore given as,
$
\Pi_i(\sigma,t)=\frac{1}{3} \epsilon_{ijk}X^j \frac{\partial X^k}{\partial \sigma}. 
$ 
Since it is expressed in terms of the coordinate variables, we
have a constrained system with three constraints:
\begin{eqnarray}
\phi_i=\Pi_i -\frac{1}{3} \epsilon_{ijk}X^j \frac{\partial X^k}{\partial \sigma} \approx 0.
\end{eqnarray}
The Poisson brackets among the constraints are given by
\begin{eqnarray}
\{\phi_i(\sigma),\phi_j(\sigma')\}=-\epsilon_{ijk}\frac{\partial X^k}{\partial \sigma}\delta(\sigma-\sigma')\,.
\end{eqnarray}
This $3\times 3$ matrix has rank two.  
It implies that a combination of the constraints $\phi_i$
is the first class. By inspection, one finds that
\begin{eqnarray}
T(\sigma)=-\frac{\partial X^i}{\partial \sigma}\phi_i
\end{eqnarray}
has vanishing bracket and becomes first class.
It satisfies a classical version of the Virasoro algebra,
\begin{eqnarray}
\left\{T(\sigma), T(\sigma')\right\}=2 T(\sigma')\partial_{\sigma'}\delta(\sigma-\sigma')
+\partial_{\sigma}T(\sigma)\delta(\sigma-\sigma')\,.
\end{eqnarray}
The appearance of the Virasoro algebra is natural
since we have the reparametrization invariance.
One may turn the first class constraints into the second class
by adding the gauge fixing condition.  There are some choices. The simplest one
is to use "static gauge",
\begin{eqnarray}\label{gf1}
\chi=X^3-\sigma\approx 0.
\end{eqnarray}
The Dirac bracket associated with it gives
\begin{eqnarray}\label{staticgauge}
\{X^1(\sigma), X^2(\sigma')\}_D=\delta(\sigma-\sigma')\,.
\end{eqnarray}
The other possibility is to use $O(3)$ invariant gauge,
\begin{eqnarray}\label{gf2}
\chi=(\partial_\sigma \vec X)^2-1\approx 0.
\end{eqnarray}
The Dirac bracket for this gauge choice gives
\begin{eqnarray}
\{X^i(\sigma),X^j(\sigma')\}_D=\epsilon_{ijk}\frac{\partial X^k}{\partial \sigma}\delta(\sigma-\sigma')\,.
\end{eqnarray}
In either case, the Nambu dynamics is described in the form of Dirac bracket as
\begin{eqnarray}
\frac{\partial X^i}{\partial t}=\left\{X^i, \omega(H,K)\right\}_D+\cdots,
\quad
\omega(H,K)=\int d\sigma H(X)\partial_\sigma K(X)\,.
\end{eqnarray}
where $\cdots$ terms are changes associated with the reparametrization of $\sigma$
to keep the consistency of gauge fixing conditions (\ref{gf1},\ref{gf2}).

This procedure seems to produce a simple 2D conformal field theory.
For example, the commutator (\ref{staticgauge}) is the same
as the commutator of $\beta-\gamma$ ghosts. A subtlety
is how to regularize the volume preserving diffeomorphism
generator $\omega(H,K)$ which are nonlinear functions of
coordinates $\vec X$.  It is also nontrivial how to recover the rotational
symmetry $O(3)$.  These issues have not been fixed in our understanding.

\section{Nambu bracket in M-theory}

In string theory,
the Lie algebra is needed when one promotes 
the low energy effective theory of a single D-brane \cite{Leigh:1989jq}
to that of a stack of multiple D-branes \cite{Witten:1995im}.
Similarly,
in M theory,
the Nambu bracket is needed
to promote the theory of a single membrance \cite{Bergshoeff:1987cm}
to multiple membranes \cite{Bagger:2006sk,Bagger:2007jr,Bagger:2007vi}.
On the other hand,
the commutator is needed for the noncommuative D-brane in the $B$-field background
\cite{Chu:1998qz,Schomerus:1999ug,Seiberg:1999vs},
and similarly the Nambu bracket is needed to formulate
an M5-brane in the $C$-field background \cite{HM,HIMS,Chen:2010br}.
\footnote{
The theory of a single M5-brane \cite{Pasti:1997gx,Bandos:1997ui,Aganagic:1997zq}
was promoted to that of multiple M5-branes in \cite{Ho:2011ni}
when they are compactified on a finite circle,
yet only the Lie bracket is used.
}
In this section, 
we review these theories of M-branes and D-branes in which 
the Nambu bracket and its generalizations appear 
to characterize the effect of interactions among branes,
or the interaction with a particular background.

\subsection{As an extension of M(atrix) theories}

The low-energy effective theories of D$p$-branes are well known to be
supersymmetric Yang-Mills theories \cite{Witten:1995im},
in which transverse coordinates $X^a$ of the target space are represented by matrices.
It was learned in the study of M(atrix) theories
that higher dimensional branes can be constructed out of lower dimensional ones
through certain matrix configurations \cite{Banks:1996nn}.
For instance, 
solutions to the Nahm equation \cite{Nahm:1979yw}
$$
\frac{dX_a}{d\sigma} + \frac{1}{2} \epsilon_{abc} [X^b, X^c] = 0
$$
for the multiple D1-brane theory
describe a bound state of D1-branes ending on a D3-brane
\cite{Diaconescu:1996rk}.
(The parameter $\sigma$ is the spatial world-sheet coordinate of the D1-brane.)
This was generalized to
the Basu-Harvey equation \cite{Basu:2004ed}
$$
\frac{dX_a}{d\sigma} + \frac{1}{6} \epsilon_{abcd} [G, X^b, X^c, X^d] = 0,
$$
to describe M2-branes ending on an M5-brane.
Here $\sigma$ is the spatial coordinate of the M2-branes
parametrizing their extension orthogonal to the M5-brane,
and $X^a$'s are the matrices representing transverse coordinates.
The 4-bracket is defined as a sum over permutations $P$ of 4 indices:
$$
[A_1, A_2, A_3, A_4] = 
\sum_{P} sgn(P) A_{P(1)}A_{P(2)}A_{P(3)}A_{P(4)}.
$$
As the matrix $G$ is fixed,
effectively a three-bracket $[G, \,\cdot\;, \,\cdot\;, \,\cdot\; ]$ appears here.
Note that a 3-bracket structure must appear 
as the M5-brane is 3-dimensional higher than an M2-brane.
Although the 3-bracket defined this way does not enjoy
enough nice algebraic properties to allow one to define
a supersymmetric action for multiple M2-branes,
this is one of the first hints that 
one should replace the Lie bracket by something like the Nambu bracket
when one considers M theory.
Another hint for the relevance of the 3-bracket to M theory 
was obtained through calculations of 
scattering amplitudes of membranes in the $C$-field background \cite{Ho:2007vk}.

As an alternative to the use of the matrix algebra to realize the Nambu bracket,
one can also define Lie 3-algebra abstractly as an analogue of the Lie algebra.
The Lie 3-algebra is defined as a linear space equipped with a totally anti-symmetrized bracket of 3 slots
$[\,\cdot\; , \,\cdot\; , \,\cdot\; ]$, 
which maps three elements to an element in the linear space.
For a given basis $\{ T^A \}$ of the linear space,
the Lie 3-bracket 
$$
[T^A, T^B, T^C] = f^{ABC}{}_D T^D
$$
is given in terms of the structure constants $f^{ABC}{}_D \in \mathbb{C}$.
The Lie 3-bracket is required to satisfy the fundamental identity
\begin{equation}
[F_1, F_2, [F_3, F_4, F_5]] =
[[F_1, F_2, F_3], F_4, F_5] + [F_3, [F_1, F_2, F_4], F_5] + [F_3, F_4, [F_1, F_2, F_5]]
\label{Fundamental-Identity}
\end{equation}
for all elements $F_1, F_2, \cdots, F_5$ of the algebra.
Lie 3-algebra is essentially the algebra of the Nambu bracket
without demanding algebraic rules of multiplication among the elements.
Hence we will refer to the Lie 3-algebra bracket also as the Nambu bracket.

A symmetric bilinear map $\langle \,\cdot\; | \,\cdot\; \rangle \in \mathbb{C}$
that maps two elements to a number
is said to be an invariant metric if we have
\begin{eqnarray}
&\langle F_1 | F_2 \rangle = \langle F_2 | F_1 \rangle, 
\\
&\langle [F_1, F_2, F_3] | F_4 \rangle + \langle F_3 | [F_1, F_2, F_4] \rangle = 0,
\label{invariant-inner-product}
\end{eqnarray}
for all elements $F_1, F_2, F_3, F_4$.


Unlike Lie algebra,
it is not clear how to realize Lie 3-algebras in terms of matrices.
Let ${\cal F}$ denote a Lie 3-algebra.
Then the Lie 3-bracket defines a set of maps
$G(F_1, F_2) \equiv [F_1, F_2, \,\cdot\;]$ as derivatives acting on ${\cal F}$ 
for every anti-symmetric pair of elements $F_1, F_2 \in {\cal F}$.
Define ${\cal G}$ to be the set of such maps;
it is obviously a Lie algebra,
of which ${\cal F}$ is a representation.
The fundamental identity (\ref{Fundamental-Identity}) implies that 
the Lie bracket of ${\cal G}$ is given by
\footnote{
This expression is not manifestly antisymmetric in 
the exchange of $(F_1, F_2)$ with $(F_3, F_4)$,
but the skew-symmetry is guaranteed by the fundamental identify
(\ref{Fundamental-Identity}).
One can thus think of Lie 3-algebras as a special class of Lie algebras
with additional internal structures.
}
$$
[G(F_1, F_2), G(F_3, F_4)]
= G([F_1, F_2, F_3], F_4) + G(F_3, [F_1, F_2, F_4]).
$$
Note that whenever there is a continuous symmetry, 
there is an associated Lie group and hence a Lie algebra.
The appearance of ${\cal G}$ and its Lie bracket is always implied by the Lie 3-algebra.

One can define gauge theories for a Lie 3-algebra ${\cal F}$
by identifying the Lie algebra ${\cal G}$ as the gauge symmetry.
For a Lie 3-algebra ${\cal F}$ with generators $\{T^A\}$,
the generators of the Lie algebra ${\cal G}$ are $\{ [ T^A, T^B, \,\cdot\; ] \}$.
A matter field $\Phi = \Phi_A T^A$ taking values in ${\cal F}$ changes by
$$
\delta \Phi = \Lambda_{AB}[T^A, T^B, \Phi]
$$
under a gauge transformation with the transformation parameters $\Lambda_{AB}$.
Equivalently,
$$
(\delta \Phi)_A = \Lambda_{CD} f^{CDB}{}_A \Phi_B = \tilde{\Lambda}^B{}_A \Phi_B,
$$
where $f^{CDB}{}_A$ is the Lie 3-algebra structure constant in the basis $\{T^A\}$,
and $\tilde{\Lambda}$ is defined by
$$
\tilde{\Lambda}^B{}_A \equiv \Lambda_{CD} f^{CDB}{}_A.
$$

The gauge potential $A_{\mu}$ takes its value in the Lie algebra ${\cal G}$:
\begin{equation}
A_{\mu} = A_{\mu AB} [T^A, T^B, \,\cdot\; ].
\label{gauge-potential}
\end{equation}
The covariant derivative $D_{\mu}$
on the base space with coordinates $\sigma^{\mu}$ is thus
$$
D_{\mu} \Phi = \frac{\partial}{\partial\sigma^{\mu}} \Phi + A_{\mu AB} [T^A, T^B, \Phi],
$$
or equivalently
$$
(D_{\mu} \Phi)_A = \frac{\partial}{\partial\sigma^{\mu}} \Phi_A + \tilde{A}_{\mu}{}^{B}{}_{A} \Phi_B,
$$
where $A_{\mu AB}$ is the gauge potential
and
$$
\tilde{A}_{\mu}{}^B{}_A \equiv A_{\mu CD} f^{CDB}{}_A.
$$
Notice that the structure constants may be such that 
a change in $A_{\mu AB}$ does not always lead to a change in $\tilde{A}_{\mu}{}^B{}_A$,
but only the components $\tilde{A}_{\mu}{}^B{}_A$ are relevant 
in the covariant derivative.

We refer to Ref.\cite{Park:2008qe} for a related idea
to use the Nambu bracket in matrix model and to Ref.\cite{Trzetrzelewski:2012re}
where it was used to describe the matrix regularization of higher dimensional spheres.

\subsection{BLG model}

The Lie 3-algebra turns out to be the appropriate symmetry structure 
for constructing a manifestly supersymmetric effective theory
for multiple M2-branes -- the Bagger-Lambert-Gustavsson (BLG) model
\footnote{
The use of the algebra with a 3-bracket is crucial for the full supersymmetry to be manifest.
An effective theory defined with the usual Lie algebra is possible \cite{Aharony:2008ug},
but only part of the supersymmetry is manifest.
}
\cite{Bagger:2006sk,Bagger:2007jr,Bagger:2007vi,Gustavsson}.

Let $x^{\mu}$ ($\mu = 0, 1, 2$) be the world-volume coordinates of M2-branes.
In addition to the gauge potential $A_{\mu}$ (\ref{gauge-potential}),
the scalar fields $X^a(x) = X^a_A(x) T^A$ ($a = 3, \cdots, 10$) 
represent the transverse coordinates,
and the 11D Majorana spinors $\Psi(x) = \Psi_A(x) T^A$ their super-partners,
which should satisfy the chirality condition 
$\Gamma_{012}\Psi = - \Psi$.
With $T_2 = 1/(2\pi \ell^3_p)$ denoting the M2-brane tension
($\ell_p$ is the M theory Planck length scale),
the action for the BLG model is \cite{Bagger:2006sk,Bagger:2007jr,Bagger:2007vi}
\begin{eqnarray}
S &=& T_2 \int d^3 x \left[
- \frac{1}{2} \langle D_{\mu}X^a | D^{\mu}X^a \rangle
- \frac{1}{12} \langle [X^a, X^b, X^c] | [X^a, X^b, X^c] \rangle
\right.
\nonumber \\
&&
+ \frac{i}{2} \langle \bar{\Psi} | \Gamma^{\mu}D_{\mu} \Psi \rangle
+ \frac{i}{4} \langle \bar{\Psi} | \Gamma_{ab} [X^a, X^b, \Psi] \rangle
\nonumber \\
&&
\left.
+ \epsilon^{\mu\nu\lambda} \left(
\frac{1}{2} f^{ABCD} A_{\mu AB} \partial_{\nu} A_{\lambda CD}
- \frac{1}{3} f^{ACD}{}_{G} g^{GH} f^{BEF}{}_{H} A_{\mu AB}A_{\nu CD}A_{\lambda EF}
\right)
\right],
\label{BLG-action}
\end{eqnarray}
where the invariant metric $g^{AB}$ is needed to define the action.

In addition to the gauge symmetry characterized by a Lie 3-algebra,
this action has the supersymmetry of 16 Grassmannian paramters.
Its SUSY transformation laws are \cite{Bagger:2006sk,Bagger:2007jr,Bagger:2007vi}
\begin{eqnarray}
\delta X^a_A &=& i\bar{\epsilon}\Gamma^a \Psi_A, \\
\delta \Psi_A &=& D_{\mu}X^a_A \Gamma^\mu\Gamma_a \epsilon  
- \frac{1}{6} X^a_B X^b_C X^c_D f^{BCD}{}_A \Gamma^{abc}\epsilon, \\
\delta \tilde{A}_{\mu}{}^B{}_A &=& 
i\bar{\epsilon}\Gamma_{\mu}\Gamma_a X^a_C \Psi_D f^{CDB}{}_A, 
\end{eqnarray}
where the SUSY transformation parameter $\epsilon$ is
an 11D Majorana spinor satisfying the chirality condition
$
\Gamma_{012}\epsilon = \epsilon
$.

A different choice of the Lie 3-algebra corresponds to a different background
for the membranes.
At the time of the proposal of the BLG model,
there were few examples of the Lie 3-algebra.
An example is the 4-generator algebra ${\cal A}_4$ 
\cite{Filippov1986,Kasymov1987}
defined by
$$
[T^A, T^B, T^C] = \epsilon_{ABCD} T^D,
$$
where $A, B, C, D = 1, 2, 3, 4$,
and the structure constant $\epsilon_{ABCD}$ is the totally anti-symmetric tensor.
The invariant metric is positive-definite and can be normalized as
$$
\langle T^A | T^B \rangle = \delta_{AB}.
$$
The algebra ${\cal A}_4$ is formally a natural generalization of the Lie algebra $su(2)$,
and the corresponding BLG model describes two M2-branes on an M-fold
\cite{Lambert:2008et,Distler:2008mk}.
More examples of Lie 3-algebras were discussed in \cite{HHM,DeMedeiros:2008zm}.

For a model to be physically interesting,
we often demand that it is free of ghosts.
Naively this seems to say that the Killing metric of the Lie 3-algebra
should be positive definite,
in order for the kinetic terms to have the correct sign for all fields.
It turns out that, however,
it is possible to define physically interesting theories 
for invariant metrics with the Lorentzian signature.

%
%
%
%

\subsubsection{BLG model for Lorentzian 3-algebra}
\hfill

\noindent
{\bf D2-branes}

It was found \cite{GMR,Benvenuti:2008bt,HIM} 
that there is a Lie 3-algebra associated with each Lie algebra,
and the BLG model defined for this Lie 3-algebra
is exactly the super Yang-Mills (SYM) action for D2-branes \cite{HIM}.
The duality between M theory and type IIA superstring theory 
is respected by the BLG model in a novel way.

Let us describe the promotion of a Lie algebra to a Lie 3-algebra
in terms of a basis of generators 
$\{ T^A \}_{A = 1}^{N}$
with the Lie bracket
$$
[T^A, T^B] = f^{AB}{}_C T^C,
$$
and the Killing form
$$
\langle T^A | T^B \rangle = h^{AB}.
$$
The associated Lie 3-algebra \cite{GMR,Benvenuti:2008bt,HIM}  can be defined 
by the following Nambu brackets:
\begin{eqnarray}
{}[T^A, T^B, T^C] &=& f^{ABC} v,
\label{3-alg-1}
\\
{}[u, T^A, T^B] &=& f^{AB}{}_C T^C,
\\
{}[v, T^A, T^B] &=& 0,
\\
{}[u, v, T^A] &=& 0,
\label{3-alg-4}
\end{eqnarray}
where $f^{ABC} \equiv f^{AB}{}_D h^{DC}$,
with two new generators $u$ and $v$.
The generator $v$ is central,
i.e., the Nambu bracket vanishes whenever it appears.
The generator $u$ has the special feature that
it never shows up on the right hand side of the Nambu bracket.
A shift of $u$ by a constant times $v$
is hence an algebra homomorphism.

The Killing form $h^{AB}$ of the Lie algebra also induces 
an invariant metric for the Lie 3-algebra:
\begin{eqnarray}
&\langle T^A | T^B \rangle = h^{AB},
\label{metric-TT}
\\
&\langle u | T^A \rangle = 0,
\qquad
\langle v | T^A \rangle = 0,
\\
&\langle u | u \rangle = 0,
\qquad
\langle v | v \rangle = 0,
\qquad
\langle u | v \rangle = 1.
\label{metric-uv}
\end{eqnarray}
As a convention,
we have normalized the metric so that $\langle u | v \rangle = 1$.
This is not the unique invariant metric,
as the requirement (\ref{invariant-inner-product})
that the inner product be invariant
allows $\langle u | u \rangle$ to be non-zero.
However, the algebra homomorphism
$$
u \rightarrow u + \alpha v 
\qquad (\alpha \in \mathbb{C})
$$
allows us to set it to zero without loss of generality.

Due to eq.(\ref{metric-uv}),
the signature of the metric is Lorentzian 
even if the Killing form $h^{AB}$ is positive definite.
As the kinetic terms of the BLG model are defined by the metric,
one should worry about the presence of negative-norm states.
The components $X_u^a, X_v^a, \Psi_u, \Psi_v$ of the matter fields
$$
X^a = X_A^a T^A + X_u^a u + X_v^a v
\qquad \mbox{and} \qquad
\Psi = \Psi_A T^A + \Psi_u u + \Psi_v v
$$
are the degrees of freedom
in danger of giving negative-norm states.
Due to the special algebraic properties of the generators $u, v$ mentioned above,
the components $X_v^a$ and $\Psi_v$ only appear as Lagrange multipliers.
The constraints they impose are free field equations for $X_u^a$ and $\Psi_u$,
although the latter also appear in the interaction terms.
A different choice of the solution of the constraints leads to
differences in the interactions,
and one obtains a slightly different model from the BLG model.
The idea of the ``Higgs mechanism'' of the BLG model
\cite{Mukhi:2008ux},
which was originally proposed for a different Lie 3-algebra ${\cal A}_4$,
suggests one to consider the special cases 
when $X_u^a, \Psi_u$ as constants
\begin{eqnarray}\label{vev}
X_u^a = 2\pi R \delta^a_{10},
\qquad
\Psi_u = 0,
\end{eqnarray}
which are solutions to the free field equations.
We have labelled the direction of the constant vector $X_u^a$
as the tenth direction in space-time without loss of generality.
It is remarkable that in this way the BLG model 
leads to exactly the super Yang-Mills theory for multiple D2-branes \cite{HIM}
obtained from compactifying M2-branes on a circle 
in the tenth direction of radius $R$.

\noindent
{\bf D$p$-branes}

The Lie 3-algebra upon which the BLG model reduces to the effective action for D2-branes
can be generalized such that the BLG Model becomes
the super Yang-Mills action for D$p$-branes for any $p \geq 2$ \cite{Ho:2009nk}.

In order to obtain the D$p$-brane action from the BLG model,
we have to enlarge the base space from $2+1$ dimensions to $p+1$ dimensions.
The additional $p-2$ coordinates $x_a$ ($a = 3, 4, \cdots, p$)
can be introduced through $p-2$ indices $\vec{m} = (m_3, m_4, \cdots, m_p)$
on the generators $T^A$, 
now denoted as $T^{\vec{m}i}$,
which can be viewed as the product of a Lie algebra generator $T^i$ 
with a function $e^{i\vec{m}\cdot\vec{x}}$ of the coordinates $\vec{x} = (x_3, x_4, \cdots, x_p)$,
and $\vec{m}$ represents the wave vector.
The Lie bracket for $T^{\vec{m}i}$ should therefore be defined by
\begin{equation}
[T^{\vec{m}i}, T^{\vec{n}j}] = f^{ij}{}_k T^{(\vec{m} + \vec{n})k}.
\label{TT=fT}
\end{equation}
In terms of this kind of Lie algebra,
in which the base-space dependence of the gauge group is 
incorporated explicitly in the Lie algebra,
one can express a $q'+1$ dimensional SYM theory 
as a $q+1$ dimensional SYM theory for any $q' > q$.
If the base space is a noncommutative space 
due to a constant $B$-field background \cite{Chu:1998qz,Schomerus:1999ug,Seiberg:1999vs},
the Lie algebra has to be a matrix algebra (e.g. $U(N)$),
and the bracket above (\ref{TT=fT}) should be changed to
\footnote{
Eqs.(32) and (34) in Ref.\cite{Ho:2009mi} are incorrect.
}
$$
[T^{\vec{m}i}, T^{\vec{n}j}] = f^{ij}{}_k \cos\left(\frac{1}{2}\theta^{ab}m_a n_b\right)T^{(\vec{m} + \vec{n})k}
+ id^{ij}{}_k \sin\left(\frac{1}{2}\theta^{ab}m_a n_b\right)T^{(\vec{m} + \vec{n})k},
$$
where $d_{ij}{}^k$ is defined by the anti-commutator
of the Lie algebra generators
$\{ T^i, T^j \} = d^{ij}{}_k T^k$.

The Lie algebra (\ref{TT=fT}) can be further extended by introducing 
generators $u_a$ corresponding to the derivatives of the coordinates $x_a$.
The Lie bracket is given by
\begin{eqnarray}
{}[u_a, u_b] &=& C_{ab} T^{\vec{0}0}, 
\\
{}[u_a, T^{\vec{m}i}] &=& m_a T^{\vec{m}i} - C_{ab} \delta^{\vec{m}}_{\vec{0}} \delta^i_0 v^b,
\\
{}[T^{\vec{m}i}, T^{\vec{n}j}] &=& m_a h^{ij} \delta^{\vec{m}+\vec{n}}_{\vec{0}} v^a + f^{ij}{}_k T^{(\vec{m}+\vec{n})k},
\\
{}[v^a, T^{\vec{m}i}] &=& 0,
\\
{}[u_a, v^b] &=& 0,
\end{eqnarray}
with constant parameters $C_{ab}$.
In the above, we have used the label $0$ for the identity matrix $T^0 = I$.
(For Lie algebras in which there is no corresponding element,
one can set it to zero in the equations above.)
The Killing form is defined as
\begin{eqnarray}
\langle u_a | v^b \rangle &=& \delta_a^b, \\
\langle T^{\vec{m}i} | T^{\vec{n}j} \rangle &=& h^{ij} \delta^{\vec{m} + \vec{n}}_{\vec{0}},
\end{eqnarray}
with all other inner products vanishing.
This is a higher loop generalization of current algebra.
As far as we know, it has never been examined in the literature
and is worth to be studied in more detail in the future.\footnote{
In mathematical literature, there is two-loop symmetry which is known
as elliptic Hall algebra $\mathfrak{gl}(1)[x^{\pm 1}, y^{\pm 1}]$
\cite{burban2005hall} which is known to be equivalent to the quantum
deformation of $W_{1+\infty}$ algebra, see
for example, \cite{miki2007q, feigin2009heisenberg, feigin2016finite}.
Since it is a quantum symmetry and not Lie algebra, it is different from
the multi-loop algebra considered here. It describes the instanton partition functions
in 5D super Yang-Mills \cite{BMZ,BFMZZ,Kimura-Pestun,Mironov:2016yue}
and the role played by the algebra seems to be similar.
}

The Lie algebra with generators $\{T^{\vec{m}i}, u_a, v^a\}$ 
can be promoted to a Lie 3-algebra in the way described above 
in eqs.(\ref{3-alg-1})--(\ref{3-alg-4})
by adjoining two more generators $u, v$.
The invariant metric can be given by (\ref{metric-TT})--(\ref{metric-uv}), too.
The BLG model with this Lie 3-algebra is then equivalent to
the super Yang-Mills theory in $p+1$ dimensions \cite{Ho:2009nk}.
The constant parameters $C_{ab}$ specify constant gauge field backgrounds.

\subsection{M5 from M2}

D$p$-branes in $B$-field background 
can be constructed out of infinitely many D$(p-2)$-branes \cite{Banks:1996nn}
(which in turn can be constructed out of lower dimensional branes
in the same fashion).
This is achieved mathematically by setting the background values of
two infinite-dimensional matrix coordinates $X_{p-1}, X_{p}$ of the D$(p-2)$-branes
to satisfy the commutation relation
$[X_{p-1}, X_p] = c I$,
where $I$ is the identity matrix and $c$ is a constant
corresponding to the gauge field background.
Similarly, 
an M5-brane in $C$-field background
can be decomposed into infinitely M2-branes \cite{HM,HIMS}.
This is achieved by using the Nambu algebra as the Lie 3-algebra in the BLG model \cite{HIM}.
Although this correspondence between M2-branes and M5-brane is expected,
mathematically it is remarkable that it can be realized explicitly for the BLG model.

In terms of a complete basis of functions $\{\chi^A(y)\}$ on a 3-manifold ${\cal M}_3$,
the Nambu bracket is 
$$
\{\chi^A, \chi^B, \chi^C\} = \frac{1}{\rho} \epsilon^{\dot{\mu}\dot{\nu}\dot{\lambda}}
\frac{\partial \chi^A}{\partial y^{\dot{\mu}}}\frac{\partial \chi^B}{\partial y^{\dot{\nu}}}\frac{\partial \chi^C}{\partial y^{\dot{\lambda}}},
$$
where $\rho$ defines the volume form $\rho dy^{\dot{1}} dy^{\dot{1}} dy^{\dot{3}}$.
We shall consider the BLG model with this algebra as the symmetry algebra,
and use the coordinates $y^{\dot{\mu}}$ with dotted indices for the internal space ${\cal M}_3$,
to be distinguished from the M2-brane world-volume coordinates $x^{\mu}$ ($\mu = 0, 1, 2$).

Since the space of functions on ${\cal M}_3$ is infinite dimensional,
the BLG model represents infinitely many M2-branes.
If a field $\Phi$ 
(e.g. $X^a(x)$ and $\Psi(x)$)
in the BLG model takes values in the Nambu algebra
$$
\Phi(x, y) = \Phi_A(x)\chi^A(y),
$$
it can be interpreted as a field living on the M5-brane world-volume,
which is the product of the 3-manifold ${\cal M}_3$ and the M2-brane world-volume.

Transformations defined by the Nambu bracket
\begin{equation}
\delta \Phi(x, y) = \Lambda_{AB}(x) \{\chi^A(y), \chi^B(y), \Phi(x, y) \}
\label{covariant-transform}
\end{equation}
is the same as a coordinate transformation in $y$,
$$
\delta \Phi = \delta y^{\dot{\mu}}(x) \partial_{\dot{\mu}} \Phi,
$$
that preserves the 3-form $\rho d^3 y$.
This 3-form $\rho d^3 y$ shall be interpreted as the $C$-field background in M theory.
Recall that a $B$-field background turns the world-volume of a D-brane into a non-commutative space 
\cite{Chu:1998qz,Schomerus:1999ug,Seiberg:1999vs},
and in the Poisson limit
the gauge symmetry on the D-brane can be identified with the diffeomorphisms
preserving the 2-form $B$-field background.
Similarly, M5-branes in $C$-field background develops the gauge symmetry of diffeomorphisms
preserving the 3-form $C$-field background.

The invariant metric can be identified with the integral
$$
\langle \chi^A | \chi^B \rangle = \int d^3 y \; \rho(y) \chi^A(y) \chi^B(y).
$$
The action of the BLG model (\ref{BLG-action}) is thus an integral over the M5-brane world-volume.

We will focus on the special case that ${\cal M}_3 = \mathbb{T}^3$,
and choose $y$ to be the Cartesian coordinates.
Then $\rho$ is just a constant,
which can be scaled to $1$ without loss of generality.

The set of functions on 3-torus $\mathbb{T}^3$ is spanned
by $\chi_{\vec n}(y)=\exp(2\pi i \vec n\cdot \vec y)$
($\vec n\in \mathbb{Z}^3$)
assuming all the radius are set to $1$ for simplicity.
In addition to them, the linear functions $u^{\dot \mu}=y^{\dot \mu}$
may enter the Nambu bracket since the derivative gives the
periodic function.  They do not show up on the right hand
side of the algebra.  In this sense, they play the role
similar to $u$ generator in (\ref{3-alg-1}).
We have to add three $v_{\dot\mu}$ generators to form
a Lorentzian triple.  As a whole, the three algebra of
Nambu-Poisson bracket is spanned by $\chi_{\vec n}$ 
($\vec n\in \mathbb{Z}^3$), $(u^{\dot\mu},v_{\dot\mu})$
and the explicit form of 3-algebra can be found in \cite{Ho:2009nk}.
We note that a similar infinite dimensional Lie 3-algebra based on Nambu bracket
was also considered in \cite{Curtright:2008jj, Curtright:2009qf}.

When we try to rewrite the BLG model in the form of a 6-dimensional field theory for the M5-brane,
it is less obvious how to replace the gauge potential 1-form $A_{\mu}$ on the M2-brane world-volume
by a 2-form gauge potential on the M5-brane.
First, 
the potential $A_{\mu}(x)$ takes values in the tensor product of the Lie 3-algebra,
so superficially it is a non-local field on the M5-brane world-volume:
$$
A_{\mu}(x, y, y') = A_{\mu AB}(x)\chi^A(y)\chi^B(y').
$$
However, 
since the gauge potential appears in the BLG model only through the form
$\tilde{A}_{\mu}{}^B{}_A \equiv A_{\mu CD} f^{CDB}{}_A$,
the BLG model only depends on $A_{\mu}$ through the local field
\begin{equation}
b_{\mu\dot{\mu}}(x, y) \equiv \left[\frac{\partial}{\partial y^{\prime\dot{\mu}}}A_{\mu}(x, y, y')\right]_{y' = y}.
\label{bdot}
\end{equation}
Hence we have some of the components of the 2-form potential derived from $A_{\mu}$.

Next we consider the scalars $X^3, X^4, X^5$ representing 
the coordinates transverse to the M2-branes but parallel to the M5-brane.
In order for the M5-brane to extend in these directions,
we choose the background values $X^3 = y^{\dot{1}}/g, X^4 = y^{\dot{2}}/g, X^5 = y^{\dot{3}}/g$
for these scalars,
where $g$ is an arbitrary constant factor of normalization.
This is parallel to (\ref{vev}).
Hence a field is defined for each of the 3 scalars as the fluctuation field:
\begin{equation}
X^{3} = \frac{y^{\dot{1}}}{g} + b^{\dot{1}}(x, y),
\qquad
X^{4} = \frac{y^{\dot{2}}}{g} + b^{\dot{2}}(x, y),
\qquad
X^{5} = \frac{y^{\dot{3}}}{g} + b^{\dot{3}}(x, y).
\end{equation}
Then we can define another set of components for the M5-brane 2-form gauge potential
\begin{equation}
b_{\dot\mu\dot\nu} \equiv \epsilon_{\dot\mu\dot\nu\dot\lambda} b^{\dot\lambda}.
\label{bdotdot}
\end{equation}

So far we have $b_{\mu\dot{\mu}}$ and $b_{\dot\mu\dot\nu}$ of the M5-brane potential,
while $b_{\mu\nu}$ is still missing.
It turns out that,
as the 3-form field strength is self-dual in the M5-brane theory,
one can formulate the gauge theory in terms of only part of the components of the gauge potential
\cite{HM,HIMS}.
A generalization of this formulation of self-dual gauge theories is available
for self-dual theories in arbitrary dimensions \cite{Chen:2010jgb}
(whenever the self-duality condition can be defined).

The covariant derivatives for this gauge symmetry can be defined as
\begin{eqnarray}
{\cal D}_{\mu}\Phi &=& \partial_{\mu}\Phi - g\{ b_{\mu\dot\mu}, y^{\dot\mu}, \Phi \},
\\
{\cal D}_{\dot\mu}\Phi &=& \frac{g^2}{2} \epsilon_{\dot\mu\dot\nu\dot\lambda}
\{ X^{\dot\nu}, X^{\dot\lambda}, \Phi \}.
\end{eqnarray}
They transform covariantly under gauge transformations 
if $\Phi$ transforms covariantly as (\ref{covariant-transform}).
It is interesting to see how the 2-form gauge potential appears in the covariant derivatives.

The field strength can be defined from 
the components (\ref{bdot}) and (\ref{bdotdot}) of the 2-form potential.
In the free field limit (or weak field limit),
they are expected to be given by
\begin{eqnarray}
{\cal H}_{\mu\dot\mu\dot\nu} &\simeq& 
\partial_{\mu}b_{\dot\mu\dot\nu} - \partial_{\dot\mu} b_{\mu\dot\nu} + \partial_{\dot\nu}b_{\mu\dot\mu} 
+ \cdots,
\\
{\cal H}_{\dot{1}\dot{2}\dot{3}} &\simeq&
\partial_{\dot{1}}b_{\dot{2}\dot{3}} + \partial_{\dot{2}}b_{\dot{3}\dot{1}} + \partial_{\dot{3}}b_{\dot{1}\dot{2}}
+ \cdots.
\end{eqnarray}
Furthermore,
they should be covariant under gauge transformations
(i.e., they transform like $\Phi$ in (\ref{covariant-transform})).
One can check that the field strength can be defined as
\begin{eqnarray}
{\cal H}_{\mu\dot\mu\dot\nu} &\equiv& 
\epsilon_{\dot\mu\dot\nu\dot\lambda} {\cal D}_{\mu}X^{\dot\lambda},
\\
{\cal H}_{\dot{1}\dot{2}\dot{3}} &\equiv&
g^2\{ X^3, X^4, X^5 \} - \frac{1}{g}.
\end{eqnarray}
For self-dual gauge theories,
the rest of the components of the field strength are redundant.

The action of the M5-brane in large $C$-field background 
derived from the BLG model this way is \cite{HIMS}
$$S = S_{B} + S_{F} + S_{CS},$$
where the bosonic part is
\begin{eqnarray}
S_B &=&
\int d^3 x d^3 y\; \Big[
- \frac{1}{2}({\cal D}_{\mu} X^a)^2 - \frac{1}{2}({\cal D}_{\dot\lambda} X^a)^2
- \frac{1}{4}{\cal H}_{\lambda\dot\mu\dot\nu}^2 - \frac{1}{12}{\cal H}_{\dot\lambda\dot\mu\dot\nu}^2
\nonumber\\
&&- \frac{g^4}{4}\{X^{\dot\mu}, X^a, X^b\}^2 - \frac{g^4}{12}\{X^a, X^b, X^c\}^2
- \frac{1}{2g^2}
\Big],
\nonumber
\end{eqnarray}
the fermionic part is
\begin{equation}
S_F = \int d^3 x d^3 y \; \Big[
\frac{i}{2} \bar\Psi\Gamma^{\mu}{\cal D}_{\mu}\Psi
+ \frac{i}{2} \bar\Psi\Gamma^{\dot\mu}{\cal D}_{\dot\mu}\Psi
+ \frac{ig^2}{2} \bar\Psi\Gamma_{\dot\mu a} \{X^{\dot\mu}, X^a, \Psi\}
- \frac{ig^2}{4} \bar\Psi\Gamma_{ab}\Gamma_{\dot{1}\dot{2}\dot{3}} \{X^a, X^b, \Psi\}
\Big]
\nonumber
\end{equation}
and the Chern-Simons part is
\begin{equation}
S_{CS} = \int d^3 x d^3 y \; \epsilon^{\mu\nu\lambda} \epsilon^{\dot\mu\dot\nu\dot\lambda}\Big[
- \frac{1}{2} \partial_{\dot\mu} b_{\mu\dot\nu} \partial_{\nu} b_{\lambda\dot\lambda}
+ \frac{g}{6} \partial_{\dot\mu} b_{\nu\dot\nu} \epsilon^{\dot\rho\dot\sigma\dot\tau}
\partial_{\dot\sigma} b_{\lambda\dot\rho} (\partial_{\dot\lambda} b_{\mu\dot\tau} - \partial_{\dot\tau} b_{\mu\dot\lambda})
\Big].
\nonumber
\end{equation}
The fermion satisfies the chirality condition
$$
\Gamma^{123}\Gamma^{\dot{1}\dot{2}\dot{3}}\Psi = - \Psi.
$$
The components $b_{\mu\nu}$ that are hidden in this formulation
can be defined when solving the field equations of this action \cite{Pasti:2009xc}.

Note that the resulting gauge theory is the first of its kind:
higher-form self-dual gauge theories with non-Abelian gauge symmetry.
The action has the correct global symmetry,
including supersymmetry, 
for an M5-brane in a large $C$-field background.
If we compactify this action on a circile in one of the $y$ directions,
we obtain the D4-brane theory in a large $B$-field background
\cite{Chen:2010br}
--- in the Poisson limit of the noncommutative gauge theory.
On the other hand,
if we compactify one of the $x$ directions,
we obtain the D4-brane theory in a large 3-form RR-field background.
Through T-dualities \cite{Ho:2011yr}, 
one can derive effective theories of D$p$-branes in NS-NS $B$-field 
or RR-field background from these D4-brane theories
\cite{Ho:2013paa}.

\noindent
{\bf D$p$-brane in R-R $(p-1)$-form field background}

While D$p$-branes in NS-NS $B$-field background are well known 
to be non-commutative gauge theories,
the effective theories for D$p$-branes in R-R $(p-1)$-form potential backgrounds 
were not known before.
What we learned from the theory of an M5-brane in the $C$-field background is that,
in addition to the usual $U(1)$ gauge symmetry for a D$p$-brane,
the R-R background turns on an additional gauge symmetry \cite{Ho:2013paa},
which is the symmetry of diffeomorphisms preserving the $(p-1)$-form background.
(Although the R-R $(p-1)$-form is not the volume form of the D$p$-brane,
we often refer to this symmetry as the volume preserving diffeomorphism.)

Under a coordinate transformation $\delta y^{\dot\mu} = \kappa^{\dot\mu}$,
a scalar field $\Phi$ transforms as
$$
\delta \Phi = \kappa^{\dot\mu}\partial_{\dot\mu}\Phi,
$$
and this transformation preserves the $(p-1)$-form $d^{p-1}y$
if $\kappa^{\dot\mu}$ is divergenceless:
$$
\partial_{\dot\mu}\kappa^{\dot\mu} = 0.
$$
Here the $y^{\dot\mu}$'s represent coordinates 
along the directions of the R-R $(p-1)$-form,
and we shall use $x^{\mu}$ ($\mu = 0, 1$) to denote the rest of the world-volume coordinates
on the D$p$-brane.

To parametrize the transformations through unconstrained functional parameters,
one can use a generalized Nambu bracket that has $(p-1)$ slots
$$
\{f_1, f_2, \cdots, f_{p-1}\} = \frac{\partial (f_1, f_2, \cdots, f_{p-1})}{\partial (y^{\dot{1}}, y^{\dot{2}}, \cdots, y^{\dot{p}-\dot{1}})}.
$$
A covariant quantity $\Phi$ transforms like
$$
\delta \Phi = \sum_{\alpha} \{f^{(\alpha)}_1, f^{(\alpha)}_2, \cdots, f^{(\alpha)}_{p-2}, \Phi\}
$$
under a gauge transformation.
Identifying the right hand side with $\kappa^{\dot\mu}\partial_{\dot\mu}\Phi$
to determine $\kappa^{\dot\mu}$,
one sees that the divergenceless condition of $\kappa^{\dot\mu}$ is automatically satisfied.

In the following we shall focus on the bosonic sector of the D$p$-brane theory
in the R-R $(p-1)$-form background
(the fermionic sector has not been worked out yet).
In the effective theory for a D$p$-brane,
the bosonic sector includes the scalars $X^a$ 
and an 1-form potential $a_{\hat{\mu}} = (a_{\mu}, a_{\dot\mu})$.
(We shall use the hatted indices $\hat{\mu}$ to refer to both the dotted ($y^{\dot\mu}$) and undotted ($x^{\mu}$) indices.)
These fields are originated from the boundary states of open strings ending on D$p$-brane \cite{Dai:1989ua}.
In the large R-R $(p-2)$-form background, 
the D$(p-2)$-branes also plays an important role,
so that by analogy (or through a series of S- and T-dualities),
there is a $(p-2)$-form potential $b_{\dot\mu_1\cdots\dot\mu_{p-2}}$
associated with the boundary states of open D$(p-2)$-branes.
This tensor field is related to the 1-form gauge potential through a duality condition
that generalizes the self-duality condition on M5-branes,
so that there is no new physical degrees of freedom on the D$p$-brane world-volume.
They play the role of the gauge potential for the gauge symmetry of volume-preserving diffeomorphisms.

It is convenient to define scalar fields $X^{\dot\mu}$ by
$$
X^{\dot\mu} \equiv \frac{y^{\dot\mu}}{g} + b^{\dot\mu}
\qquad
\left(
b^{\dot\mu} \equiv \frac{1}{(p-2)!}\epsilon^{\dot\mu\dot\mu_1\cdots\dot\mu_{p-2}}b_{\dot\mu_1\cdots\dot\mu_{p-2}}
\right),
$$
so that the gauge transformation property of the gauge field $b_{\dot\mu_1\cdots\dot\mu_{p-2}}$
is equivalent to the condition that $X^{\dot\mu}$ transform covariantly.
While both $X^{\dot\mu}$ and $X^a$ transform covariantly,
the 1-form potential transforms by
$$
\delta a_{\hat{\mu}} = \partial_{\hat{\mu}} \lambda 
+ g(\kappa^{\dot{\nu}}\partial_{\dot{\nu}}a_{\hat{\mu}} + a_{\dot\nu}\partial_{\hat{\mu}}\kappa^{\dot\nu}),
$$
where the first term is the usual $U(1)$ gauge transformation.

In terms of the following definitions
\begin{eqnarray}
F_{\hat{\mu}\hat{\nu}} &\equiv& \partial_{\hat{\mu}}a_{\hat{\nu}} - \partial_{\hat{\nu}}a_{\hat{\mu}},
\\
V_{\dot\nu}{}^{\dot\mu} &\equiv& \delta_{\dot\nu}^{\dot\mu} + g \partial_{\dot\nu}b^{\dot\mu},
\\
M_{\dot\mu\dot\nu}{}^{\mu\nu} &\equiv&
V_{\dot\mu}{}^{\dot\lambda}V_{\dot\nu\dot\lambda}\delta^{\mu\nu}
- g\epsilon^{\mu\nu}F_{\dot\mu\dot\nu},
\\
\hat{B}_{\mu}{}^{\dot\mu} &\equiv&
(M^{-1})_{\mu\nu}{}^{\dot\mu\dot\nu}
(V_{\dot\nu}{}^{\dot\lambda}\partial^{\nu}b_{\dot\lambda} + \epsilon^{\nu\lambda}F_{\lambda\dot\nu}
+ g\partial_{\dot\nu}X^a \partial^{\nu}X^a),
\end{eqnarray}
where $M^{-1}$ is defined by
$M_{\dot\mu\dot\nu}{}^{\mu\nu}M^{-1}{}_{\nu\lambda}{}^{\dot\nu\dot\lambda} = \delta_{\dot\mu}^{\dot\lambda}\delta_{\lambda}^{\mu}$,
the covariant derivatives of a covariant field $\Phi$ are defined as
\begin{eqnarray}
{\cal D}_{\mu} \Phi &=& \partial_{\mu}\Phi - g\hat{B}_{\mu}{}^{\dot\mu}\partial_{\dot\mu}\Phi,
\\
{\cal D}_{\dot\mu} \Phi &=& \frac{(-1)^p}{(p-2)!}g^{p-2}
\epsilon_{\dot\mu\dot\mu_1\cdots\dot\mu_{p-2}}\{X^{\dot\mu_1}, \cdots, X^{\dot\mu_{p-2}}, \Phi\}.
\end{eqnarray}

The usual definition of the Abelian field strength $F_{\hat{\mu}\hat{\nu}}$ 
is no longer suitable as it is not covariant.
Proper definitions of the field strength for $a_{\hat{\mu}}$ are
\begin{eqnarray}
{\cal F}_{\dot\mu\dot\nu} &=& \frac{g^{p-3}}{(p-3)!}
\epsilon_{\dot\mu\dot\nu\dot\mu_1\cdots\dot\mu_{p-3}}
\{X^{\dot\mu_1}, \cdots, X^{\dot\mu_{p-3}}, a_{\dot\nu}, y^{\dot\nu}\},
\\
{\cal F}_{\mu\dot\mu} &=& V^{-1}_{\dot\mu}{}^{\dot\nu}
(F_{\mu\dot\nu} + gF_{\dot\nu\dot\lambda} \hat{B}_{\mu}{}^{\dot\lambda}),
\\
{\cal F}_{\mu\nu} &=& F_{\mu\nu} + g[-F_{\mu\dot\mu}\hat{B}_{\nu}{}^{\dot\mu}
+ F_{\nu\dot\mu}\hat{B}_{\mu}{}^{\dot\mu} + gF_{\dot\mu\dot\nu}\hat{B}_{\mu}{}^{\dot\mu}\hat{B}_{\nu}{}^{\dot\nu}].
\end{eqnarray}
On the other hand,
the field strength for $b_{\dot\mu_1\cdots\dot\mu_{p-2}}$,
\begin{eqnarray}
{\cal H}^{\dot\mu_1\cdots\dot\mu_{p-1}} &\equiv& 
g^{p-2} \{ X^{\dot\mu_1}, \cdots, X^{\dot\mu_{p-1}} \}
- \frac{1}{g} \epsilon^{\dot\mu_1\cdots\dot\mu_{p-1}}
\end{eqnarray}
is (up to a constant) just one of a class of covariant quantities defined by
\begin{eqnarray}
{\cal O}^{\dot\mu_1\cdots\dot\mu_{\ell} a_1\cdots a_{p-1-\ell-2m}}_{(\ell m)}
&=&
\{ X^{\dot\mu_1}, \cdots, X^{\dot\mu_{\ell}}, a_{\dot\nu_1}, \cdots, a_{\dot\nu_m}, 
\frac{y^{\dot\nu_1}}{g}, \cdots, \frac{y^{\dot\nu_m}}{g}, X^{a_1}, \cdots, X^{a_{p-1-\ell-2m}} \},
\end{eqnarray}
where $\ell$, $m$ are arbitrary non-negative integers such that $\ell + 2m \leq p-1$.

The bosonic part of the action is found to be given by
\begin{eqnarray}
S_{Dp} &=& \int d^{p+1} x\; \Big[
- \frac{1}{2} ({\cal D}_{\mu}X^a)^2 + \frac{1}{2g} \epsilon^{\mu\nu}{\cal F}_{\mu\nu}
+ \frac{1}{2} {\cal F}_{\mu\dot\mu}^2 
\nonumber \\
&& - \frac{g^{2(p-2)}}{2} \sum_{(\ell, m) \in S} \frac{1}{\ell ! (m!)^2 (q-\ell-2m)!}  {\cal O}_{\ell m}^2
\Big],
\end{eqnarray}
where
$$
{\cal O}^2_{\ell m} \equiv \{ X^{\dot\mu_1}, \cdots, X^{\dot\mu_{\ell}}, a_{\dot\nu_1}, \cdots, a_{\dot\nu_m}, 
\frac{y^{\dot\nu_1}}{g}, \cdots, \frac{y^{\dot\nu_m}}{g}, X^{a_1}, \cdots, X^{a_{p-1-\ell-2m}} \}^2
$$
and
$$
S = \{ (\ell, m) | \ell, m \geq 0; \ell + 2m \leq p-1 \}.
$$
This result allows one to check explicitly the S-duality for D3-branes 
in the NS-NS and R-R field background \cite{Ho:2013opa}.
Unlike the case of trivial background,
where the S-duality is a quantum theory that cannot be verified directly by field redefinitions,
the D3-brane in large NS-NS and R-R 2-form backgrounds can be explicitly verified.

\section{Conclusion}

The Nambu bracket was first proposed as a generalization 
of the Poisson bracket for the canonical formulation of physical systems.
In particular,
the Nambu bracket and its generalizations found its natural applications
to systems involving extended objects.

One may wonder whether the use of Nambu bracket is unavoidable,
or how much advantage it can bring to us.
On this aspect, 
we recall that in the canonical formulation,
the Poisson bracket cannot be fixed without a complete gauge fixing
when there is gauge symmetry.
The definition of the Poisson bracket depends on the choice of gauge.
On the other hand,
it was shown \cite{Chu:2010eb} that,
in certain examples,
a Nambu bracket can be defined without gauge fixing,
such that when a gauge-fixing condition $f=0$ is chosen,
the Poisson bracket $\{\cdot, \cdot\}_f$ for that gauge is simply given by
$$
\{A, B\}_f = \{A, B, f\},
$$
for any choice of gauge $f$.
It is therefore a generalization of the canonical formulation 
that is gauge-independent.
This trick can be extended to a generic constrained system \cite{curtright2003quantizing, Horikoshi:2013yxa, HM-future}.
In general, 
a constrained system with $N$ constraints can be 
formulated with a generalized Nambu bracket with $N+2$ slots.

Like the Poisson bracket, 
the Nambu bracket and its generalizations 
also found their use in describing symmetries and interactions
for various systems,
including vortices and branes.
The Nambu bracket is used in the description of
a system of multiple M2-branes
and a single M5-brane in $C$-field background.
A $(p-1)$-bracket is used in the theory of
a single D$p$-brane in the R-R $(p-1)$-form background.

The quantization of the Nambu bracket remains elusive. 
People have tried using matrices and even nonassociative algebras
to define Nambu brackets,
but it seems hard to satisfy the fundamental identity,
at least not in the same fashion that 
the Jacobi identity is satisfied by the commutator of an associative algebra.
The Zariski algebra provides a quantization of the Nambu algebra,
but it is unclear how it can be applied in a physical theory 
as a small deformation of the classical Nambu algebra.
For instance,
the theory of a single M5-brane in $C$-field background
involves the Nambu bracket.
\footnote{
To emphasize that this Nambu bracket is classical,
i.e., before quantization, 
the classical Nambu bracket is often referred to as the Nambu-Poisson bracket,
and hence this M5-brane theory is referred to as the NP M5-brane.
}
Upon double dimension reduction,
it reduces to the Poisson limit of the noncommutative D4-brane.
One would like to deform the Nambu-Poisson algebra in the M5-brane theory 
such that the double dimension reduction leads to 
the full noncommutative D4-brane.
But there is a no-go theorem \cite{Chen:2010ny} against this possibility.

In the case of D-branes,
a single D-brane in $B$-field background 
and a multiple D-brane system share the same 
algebraic structure of non-Abelian gauge symmetry
characterized by the definition of commutators.
This leads us to suspect that
if one can quantize the Nambu-Poisson bracket,
it would perhaps lead us to the mysterious non-Abelian gauge symmetry of multiple M5-branes.
Over 40 years after Nambu's introduction,
reviewing the fruitful results inspired by the idea of the Nambu bracket,
we believe that there are still much more remarkable results to come
related to the Nambu bracket.

\section*{Acknowledgment}

YM would like to thank the organizers of Nambu memorial symposium
to provide an opportunity to give a talk on the Nambu bracket.
He is obliged to Prof. Nambu for his hospitality during
his stay at the University of Chicago as a postdoc fellow in 1989.
The discussions and the conversations with Prof. Nambu
have been invaluable experience for him. He is partially supported by Grants-in-Aid for Scientific Research (Kakenhi \#25400246) from MEXT, Japan.
The work of PMH is supported in part by the Ministry of Science and Technology, R.O.C.,
and by National Taiwan University.

\bibliography{HM_0331}

\begin{thebibliography}{10}

\bibitem{Nambu:1973qe}
Yoichiro Nambu, Phys.Rev., {\bf D7}, 2405--2412 (1973).

\bibitem{Takhtajan:1993vr}
Leon Takhtajan, Commun. Math. Phys., {\bf 160}, 295--316 (1994),
  {{arXiv:hep-th/9301111}}.

\bibitem{Bagger:2006sk}
Jonathan Bagger and Neil Lambert, Phys. Rev., {\bf D75}, 045020 (2007),
  {{arXiv:hep-th/0611108}}.

\bibitem{Bagger:2007jr}
Jonathan Bagger and Neil Lambert, Phys. Rev., {\bf D77}, 065008 (2008),
  {{arXiv:0711.0955}}.

\bibitem{Bagger:2007vi}
Jonathan Bagger and Neil Lambert, JHEP, {\bf 02}, 105 (2008),
  {{arXiv:0712.3738}}.

\bibitem{Gustavsson}
Andreas Gustavsson, Nucl. Phys., {\bf B811}, 66--76 (2009),
  {{arXiv:0709.1260}}.

\bibitem{Curtright:2002sr}
Thomas~L. Curtright and Cosmas~K. Zachos, New J. Phys., {\bf 4}, 83 (2002),
  {{arXiv:hep-th/0205063}}.

\bibitem{gautheron1996some}
Philippe Gautheron, Letters in Mathematical Physics, {\bf 37}(1), 103--116
  (1996).

\bibitem{Cherkis:2014xua}
Sergey~A. Cherkis, Lett. Math. Phys., {\bf 105}(5), 641--659 (2015),
  {{arXiv:1403.6836}}.

\bibitem{Dito:1996xr}
Giuseppe Dito, Moshe Flato, Daniel Sternheimer, and Leon Takhtajan, Commun.
  Math. Phys., {\bf 183}, 1--22 (1997),  {{arXiv:hep-th/9602016}}.

\bibitem{Minic:1999js}
D.~Minic (1999),  {{arXiv:hep-th/9909022}}.

\bibitem{Curtright:2002fd}
Thomas Curtright and Cosmas~K. Zachos, Phys. Rev., {\bf D68}, 085001 (2003),
  {{arXiv:hep-th/0212267}}.

\bibitem{Banks:1996vh}
Tom Banks, W.~Fischler, S.~H. Shenker, and Leonard Susskind, Phys. Rev., {\bf
  D55}, 5112--5128 (1997),  {{arXiv:hep-th/9610043}}.

\bibitem{Awata:1999dz}
Hidetoshi Awata, Miao Li, Djordje Minic, and Tamiaki Yoneya, JHEP, {\bf 02},
  013 (2001),  {{arXiv:hep-th/9906248}}.

\bibitem{Yoneya:2016wqw}
Tamiaki Yoneya (2016),  {{arXiv:1603.06402}}.

\bibitem{Kawamura:2002yz}
Yoshiharu Kawamura, Prog. Theor. Phys., {\bf 109}, 153--168 (2003),
  {{arXiv:hep-th/0207054}}.

\bibitem{Kawamura:2003cw}
Yoshiharu Kawamura, Prog. Theor. Phys., {\bf 110}, 579--587 (2003),
  {{arXiv:hep-th/0304149}}.

\bibitem{Ho:2007vk}
Pei-Ming Ho and Yutaka Matsuo, Gen.Rel.Grav., {\bf 39}, 913--944 (2007),
  {{arXiv:hep-th/0701130}}.

\bibitem{Turaev:1992hq}
V.~G. Turaev and O.~Y. Viro, Topology, {\bf 31}, 865--902 (1992).

\bibitem{Bergshoeff:2000jn}
E.~Bergshoeff, D.~S. Berman, J.~P. van~der Schaar, and P.~Sundell, Nucl. Phys.,
  {\bf B590}, 173--197 (2000),  {{arXiv:hep-th/0005026}}.

\bibitem{Kawamoto:2000zt}
Shoichi Kawamoto and Naoki Sasakura, JHEP, {\bf 07}, 014 (2000),
  {{arXiv:hep-th/0005123}}.

\bibitem{Saitou:2014vwa}
Mayumi Saitou, Kazuharu Bamba, and Akio Sugamoto, PTEP, {\bf 2014}, 103B03
  (2014),  {{arXiv:1408.3885}}.

\bibitem{Lund:1976ze}
Fernando Lund and Tullio Regge, Phys. Rev., {\bf D14}, 1524 (1976).

\bibitem{Matsuo:1993ie}
Yutaka Matsuo, Mod. Phys. Lett., {\bf A8}, 2677--2686 (1993),
  {{arXiv:hep-th/9305151}}.

\bibitem{salmon2007general}
Rick Salmon, Journal of the Atmospheric Sciences, {\bf 64}(2), 515--531 (2007).

\bibitem{nevir2009energy}
Peter N{\'e}vir and Matthias Sommer, Journal of the Atmospheric Sciences, {\bf
  66}(7), 2073 (2009).

\bibitem{sommer2009conservative}
Matthias Sommer and Peter N{\'e}vir, Quarterly Journal of the Royal
  Meteorological Society, {\bf 135}(639), 485--494 (2009).

\bibitem{Pioline:2002ba}
B.~Pioline, Phys. Rev., {\bf D66}, 025010 (2002),  {{arXiv:hep-th/0201257}}.

\bibitem{Matsuo:2000fh}
Y.~Matsuo and Y.~Shibusa, JHEP, {\bf 02}, 006 (2001),
  {{arXiv:hep-th/0010040}}.

\bibitem{Leigh:1989jq}
R.G. Leigh, Modern Physics Letters A, {\bf 04}(28), 2767--2772 (dec 1989).

\bibitem{Witten:1995im}
Edward Witten, Nucl. Phys., {\bf B460}, 335--350 (1996),
  {{arXiv:hep-th/9510135}}.

\bibitem{Bergshoeff:1987cm}
E.~Bergshoeff, E.~Sezgin, and P.~K. Townsend, Phys. Lett., {\bf B189}, 75--78
  (1987).

\bibitem{Chu:1998qz}
Chong-Sun Chu and Pei-Ming Ho, Nucl. Phys., {\bf B550}, 151--168 (1999),
  {{arXiv:hep-th/9812219}}.

\bibitem{Schomerus:1999ug}
Volker Schomerus, JHEP, {\bf 06}, 030 (1999),  {{arXiv:hep-th/9903205}}.

\bibitem{Seiberg:1999vs}
Nathan Seiberg and Edward Witten, JHEP, {\bf 09}, 032 (1999),
  {{arXiv:hep-th/9908142}}.

\bibitem{HM}
Pei-Ming Ho and Yutaka Matsuo, JHEP, {\bf 0806}, 105 (2008),
  {{arXiv:0804.3629}}.

\bibitem{HIMS}
Pei-Ming Ho, Yosuke Imamura, Yutaka Matsuo, and Shotaro Shiba, JHEP, {\bf 08},
  014 (2008),  {{arXiv:0805.2898}}.

\bibitem{Chen:2010br}
Chien-Ho Chen, Kazuyuki Furuuchi, Pei-Ming Ho, and Tomohisa Takimi, JHEP, {\bf
  10}, 100 (2010),  {{arXiv:1006.5291}}.

\bibitem{Pasti:1997gx}
Paolo Pasti, Dmitri~P. Sorokin, and Mario Tonin, Phys. Lett., {\bf B398},
  41--46 (1997),  {{arXiv:hep-th/9701037}}.

\bibitem{Bandos:1997ui}
Igor~A. Bandos, Kurt Lechner, Alexei Nurmagambetov, Paolo Pasti, Dmitri~P.
  Sorokin, and Mario Tonin, Phys. Rev. Lett., {\bf 78}, 4332--4334 (1997),
  {{arXiv:hep-th/9701149}}.

\bibitem{Aganagic:1997zq}
Mina Aganagic, Jaemo Park, Costin Popescu, and John~H. Schwarz, Nucl. Phys.,
  {\bf B496}, 191--214 (1997),  {{arXiv:hep-th/9701166}}.

\bibitem{Ho:2011ni}
Pei-Ming Ho, Kuo-Wei Huang, and Yutaka Matsuo, JHEP, {\bf 07}, 021 (2011),
  {{arXiv:1104.4040}}.

\bibitem{Banks:1996nn}
Tom Banks, Nathan Seiberg, and Stephen~H. Shenker, Nucl. Phys., {\bf B490},
  91--106 (1997),  {{arXiv:hep-th/9612157}}.

\bibitem{Nahm:1979yw}
W.~Nahm, Phys. Lett., {\bf B90}, 413 (1980).

\bibitem{Diaconescu:1996rk}
Duiliu-Emanuel Diaconescu, Nucl. Phys., {\bf B503}, 220--238 (1997),
  {{arXiv:hep-th/9608163}}.

\bibitem{Basu:2004ed}
Anirban Basu and Jeffrey~A. Harvey, Nucl. Phys., {\bf B713}, 136--150 (2005),
  {{arXiv:hep-th/0412310}}.

\bibitem{Park:2008qe}
Jeong-Hyuck Park and Corneliu Sochichiu, Eur. Phys. J., {\bf C64}, 161--166
  (2009),  {{arXiv:0806.0335}}.

\bibitem{Trzetrzelewski:2012re}
Maciej Trzetrzelewski, Nucl. Phys., {\bf B864}, 869--883 (2012),
  {{arXiv:1206.7060}}.

\bibitem{Aharony:2008ug}
Ofer Aharony, Oren Bergman, Daniel~Louis Jafferis, and Juan Maldacena, JHEP,
  {\bf 10}, 091 (2008),  {{arXiv:0806.1218}}.

\bibitem{Filippov1986}
V.~T. Filippov, Siberian Mathematical Journal, {\bf 26}(6), 879--891 (1986).

\bibitem{Kasymov1987}
Sh.~M. Kasymov, Algebra and Logic, {\bf 26}(3), 155--166 (jun 1987).

\bibitem{Lambert:2008et}
Neil Lambert and David Tong, Phys. Rev. Lett., {\bf 101}, 041602 (2008),
  {{arXiv:0804.1114}}.

\bibitem{Distler:2008mk}
Jacques Distler, Sunil Mukhi, Constantinos Papageorgakis, and Mark
  Van~Raamsdonk, JHEP, {\bf 05}, 038 (2008),  {{arXiv:0804.1256}}.

\bibitem{HHM}
Pei-Ming Ho, Ru-Chuen Hou, and Yutaka Matsuo, JHEP, {\bf 06}, 020 (2008),
  {{arXiv:0804.2110}}.

\bibitem{DeMedeiros:2008zm}
Paul De~Medeiros, Jose~M. Figueroa-O'Farrill, and Elena Mendez-Escobar, JHEP,
  {\bf 07}, 111 (2008),  {{arXiv:0805.4363}}.

\bibitem{GMR}
Jaume Gomis, Giuseppe Milanesi, and Jorge~G. Russo, JHEP, {\bf 06}, 075 (2008),
   {{arXiv:0805.1012}}.

\bibitem{Benvenuti:2008bt}
Sergio Benvenuti, Diego Rodriguez-Gomez, Erik Tonni, and Herman Verlinde, JHEP,
  {\bf 01}, 078 (2009),  {{arXiv:0805.1087}}.

\bibitem{HIM}
Pei-Ming Ho, Yosuke Imamura, and Yutaka Matsuo, JHEP, {\bf 07}, 003 (2008),
  {{arXiv:0805.1202}}.

\bibitem{Mukhi:2008ux}
Sunil Mukhi and Constantinos Papageorgakis, JHEP, {\bf 05}, 085 (2008),
  {{arXiv:0803.3218}}.

\bibitem{Ho:2009nk}
Pei-Ming Ho, Yutaka Matsuo, and Shotaro Shiba, JHEP, {\bf 0903}, 045 (2009),
  {{arXiv:0901.2003}}.

\bibitem{Ho:2009mi}
Pei-Ming Ho, Nucl. Phys., {\bf A844}, 95C--108C (2010),  {{arXiv:0912.0055}}.

\bibitem{burban2005hall}
Igor Burban and Olivier Schiffmann, arXiv preprint math/0505148 (2005).

\bibitem{miki2007q}
Kei Miki, Journal of Mathematical Physics, {\bf 48}(12), 3520 (2007).

\bibitem{feigin2009heisenberg}
Boris Feigin and Alexander Tsymbaliuk, arXiv preprint arXiv:0904.1679 (2009).

\bibitem{feigin2016finite}
B~Feigin, M~Jimbo, T~Miwa, and E~Mukhin, arXiv preprint arXiv:1603.02765
  (2016).

\bibitem{BMZ}
Jean-Emile Bourgine, Yutaka Matsuo, and Hong Zhang, arXiv preprint
  arXiv:1512.02492 (2015).

\bibitem{BFMZZ}
J.-E. Bourgine, M.~Fukuda, Y.~Matsuo, R.-D. Zhu, and H.~Zhang, to appear
  (2016).

\bibitem{Kimura-Pestun}
Taro Kimura and Vasily Pestun, arXiv preprint arXiv:1512.08533 (2015).

\bibitem{Mironov:2016yue}
A.~Mironov, A.~Morozov, and Y.~Zenkevich (2016),  {{arXiv:1603.05467}}.

\bibitem{Curtright:2008jj}
Thomas~L. Curtright, David~B. Fairlie, and Cosmas~K. Zachos, Phys. Lett., {\bf
  B666}, 386--390 (2008),  {{arXiv:0806.3515}}.

\bibitem{Curtright:2009qf}
Thomas Curtright, Xiang Jin, Luca Mezincescu, David Fairlie, and Cosmas~K.
  Zachos, Phys. Lett., {\bf B675}, 387--392 (2009),  {{arXiv:0903.4889}}.

\bibitem{Chen:2010jgb}
Wei-Ming Chen and Pei-Ming Ho, Nucl. Phys., {\bf B837}, 1--21 (2010),
  {{arXiv:1001.3608}}.

\bibitem{Pasti:2009xc}
Paolo Pasti, Igor Samsonov, Dmitri Sorokin, and Mario Tonin, Phys. Rev., {\bf
  D80}, 086008 (2009),  {{arXiv:0907.4596}}.

\bibitem{Ho:2011yr}
Pei-Ming Ho and Chi-Hsien Yeh, JHEP, {\bf 03}, 143 (2011),
  {{arXiv:1101.4054}}.

\bibitem{Ho:2013paa}
Pei-Ming Ho and Chen-Te Ma, JHEP, {\bf 05}, 056 (2013),  {{arXiv:1302.6919}}.

\bibitem{Dai:1989ua}
Jin Dai, R.G. Leigh, and Joseph Polchinski, Modern Physics Letters A, {\bf
  04}(21), 2073--2083 (oct 1989).

\bibitem{Ho:2013opa}
Pei-Ming Ho and Chen-Te Ma, JHEP, {\bf 11}, 142 (2014),  {{arXiv:1311.3393}}.

\bibitem{Chu:2010eb}
Chong-Sun Chu and Pei-Ming Ho, JHEP, {\bf 02}, 020 (2011),
  {{arXiv:1011.3765}}.

\bibitem{curtright2003quantizing}
Thomas Curtright and Cosmas Zachos, AIP Conf. Proc., {\bf 672}, 165 (2003),
  {{hep-th/0303088}}.

\bibitem{Horikoshi:2013yxa}
Atsushi Horikoshi and Yoshiharu Kawamura, PTEP, {\bf 2013}(7), 073A01 (2013),
  {{arXiv:1304.2086}}.

\bibitem{HM-future}
Pei-Ming Ho and Yutaka Matsuo, to appear (2016).

\bibitem{Chen:2010ny}
Chien-Ho Chen, Pei-Ming Ho, and Tomohisa Takimi, JHEP, {\bf 03}, 104 (2010),
  {{arXiv:1001.3244}}.

\end{thebibliography}

\end{document}